%% file: GMM_outline.tex
\nonstopmode

\documentclass[11pt]{article}
\usepackage{natbib}
\usepackage{url}
\usepackage{fancyhdr}
\usepackage{float}
\usepackage{booktabs}
 \usepackage[utf8]{inputenc}

\usepackage{dcolumn}

\usepackage{amsthm}
\usepackage{amsmath}
\usepackage{amsbsy}
\usepackage{amsfonts}
\usepackage{bbold}
\usepackage{xfrac}

\usepackage[normalem]{ulem}
\useunder{\uline}{\ul}{}
\usepackage{caption}
\usepackage{graphicx}
\usepackage{color}
\DeclareMathSymbol{\real}{\mathord}{AMSb}{"52}
\DeclareMathSymbol{\natural}{\mathord}{AMSb}{"4E}
\DeclareMathSymbol{\prob}{\mathord}{AMSb}{"50}
\DeclareMathSymbol{\blackbox}{\mathord}{AMSa}{"04}

\usepackage{amsmath}
\usepackage{accents}
\newlength{\dhatheight}

\usepackage[pdftex]{hyperref}

\usepackage{xcolor}
\usepackage{mdframed}

\newtheorem{proposition}{Proposition}

\newtheorem{corollary}{Corollary}

\newenvironment{definition}{\vspace{.4cm} \noindent {\bf Definition:}}{\vspace{.4cm}}
\newmdenv[
  backgroundcolor=white,
  linecolor=black,
  linewidth=2pt,
  roundcorner=5pt,
  innertopmargin=10pt,
  innerbottommargin=10pt,
  innerleftmargin=10pt,
  innerrightmargin=10pt,
  font=\color{black},
  frametitlefont=\color{black}\bfseries,
  frametitlebackgroundcolor=white
]{exampleboxinner}

\newenvironment{examplebox}[1]
{\begin{exampleboxinner}[frametitle={#1}]}
{\end{exampleboxinner}}

\begin{document}
\title{Pooling Liquidity Pools in AMMs}
%
%

\author{
    Marcelo Bagnulo, Angel Hernando-Veciana, Efthymios Smyrniotis\thanks{This research is part of the project I+D+i TED2021-131844B-I00, funded by MCIN/ AEI/10.13039/501100011033 and the European Union NextGeneration EU/PRTR. This work was partially funded by the European Union through NGI Sargasso’s GMM project under Horizon Europe (Grant Agreement No. 101092887). We are grateful to Francisco Peñaranda and Jonathan Jiménez Muñoz for their invaluable support with the data analysis.} \\
    \small Univ. Carlos III de Madrid
}

%
%
\date{May, 2025}

\maketitle              

\begin{abstract}
Market fragmentation across multiple Automated Market Makers (AMMs) creates inefficiencies such as costly arbitrage, unnecessarily high slippage and delayed incorporation of new information into prices. These inefficiencies raise trading costs, reduce liquidity provider profits, and degrade overall market efficiency. To address these issues, we propose a modification of the Constant Product Market Maker (CPMM) pricing mechanism, called the Global Market Maker (GMM), which aggregates liquidity information from all AMMs to mitigate these inefficiencies. Through theoretical and numerical analyses, we demonstrate that the GMM enhances profits for both AMMs and traders by eliminating arbitrage opportunities, and reducing the profitability of sandwich attacks and impermanent losses.

\vspace{.3cm}

{\bf Keywords:}
Decentralized Exchanges (DEXs) \and Automated Market Makers (AMMs) \and Constant Product Market Maker (CPMM)  \and Arbitrage \and Slippage \and Sandwich Attacks \and Impermanent Loss. 

\end{abstract}
\newpage 

\section{Introduction}

\input{sections/intro}

\section{Literature Review}

This paper is related to a relatively new but fast-growing literature on formalizing and studying the optimal design of constant function market makers (CFMMs) in decentralized exchanges. \cite{angeris2020cfmm}  introduce a general  framework for CFMMs, proving their efficiency in aggregating information and setting prices through  reserve functions, in the presence of arbitrage. \cite{angeris2021uniswap} provide a detailed analysis of Uniswap’s AMM, showing how liquidity depth, trade size, fees, and arbitrage activity, interact to determine price slippage, efficiency, and risks for liquidity providers. \cite{milionis2022optimal}  propose optimal AMM designs tailored to different market goals, such as incentivizing liquidity provision. \cite{aoyagi2023amms} develops a game-theoretic model of AMM markets, analyzing how equilibrium  and trader surplus depend on AMM parameters and compare these to traditional double auction markets. Similarly, \cite{milionis2023dex}  study equilibrium and efficiency in AMM-based decentralized exchanges, deriving conditions for price efficiency and welfare, comparing AMM with order book models. \cite{evans2021option} linked AMM mechanisms to classical option pricing, revealing that liquidity providers face similar risks to options writers. \cite{Lehar2021}  provide   theoretical and empirical analysis of Uniswap’s AMM, modeling equilibrium pool sizes, liquidity provision, and the role of arbitrageurs in  short-run price discovery. They empirically show that AMMs can offer stable liquidity and efficient price alignment with centralized exchanges. Our paper contributes to this literature by proposing a design that incorporates market information to tackle fragmentation and the need for arbitrageurs, while increasing the incentives for liquidity provision and trade.  

This paper is also related to the literature that documents and measures inefficiencies that characterize CFMMs, namely arbitrage and impermanent loss for liquidity providers, and MEV operations for traders. \cite{milionis2024automatedmarketmakinglossversusrebalancing} identifies arbitrage as a rebalancing cost for liquidity providers in a theoretical model of liquidity provision. \cite{milionis2023automatedmarketmakingarbitrage} study the effect of fees on arbitrage. \cite{Mazor2023} explores cross-chain arbitrage in decentralized exchanges, providing an empirical analysis beyond the Ethereum ecosystem. \cite{hansson2022arbitrage} examine arbitrage opportunities in Ethereum, identifying \$30 million in cross-exchange and triangular arbitrage profits across 63,168 trading pairs between July 2020 and February 2022.  \cite{daian2019flashboys20frontrunning} introduce the concept of MEV.\cite{Canidio2024FrontRunning} provide a game-theoretic analysis of sandwich attacks. \cite{qin2021quantifyingblockchainextractablevalue} estimate that total extractable value over a 32-month period reached \$541 million, highlighting the scale of MEV extraction. \cite{chi2024remeasuringarbitragesandwichattacks} develop methodologies to identify sandwich attacks, finding that \$675 million was extracted before September 2022 and noting that high-volatility tokens are primary MEV targets. \cite{Gramlich2020} provide a comprehensive literature review on MEV, categorizing extraction strategies and countermeasures. \cite{Weintraub2022} analyze MEV extraction in private pools, quantifying the role of flashbots and other sophisticated arbitrage techniques. Moreover, (\cite{evans2021option}; \cite{loesch2021automated};\cite{cartea2024microstructure}) find that AMM liquidity providers frequently incur impermanent loss that is not fully offset by fee income, particularly in volatile markets. Lastly, \cite{cartea2023arbitrage} derive the optimal strategy for arbitrageurs. Our design improves on all those fronts and we quantify the improvement using simulations from different AMMs and time periods. 

Lastly, our paper contributes to the literature that explores AMM designs to overcome the aforementioned issues. \cite{cartea2023strategic} propose a general framework that allows liquidity providers to strategically change the quoted marginal price and the impact of an order to that price, maximizing their profits. Their framework allows for the incorporation of information outside the market but do not address potential issues with oracles. \footnote{For an analysis on the manipulation of oracles see \cite{Caldarelli}.} Our model even though less general, allows us to  explicitly study ways to incorporate market information, avoiding manipulation, and to study the effects of the proposed mechanism to both liquidity providers and traders.  \cite{canidio2024arbitrageurs} introduce batch trading, a system where orders are executed in batches rather than continuously, thereby reducing arbitrage and sandwich attacks. This mechanism leverages arbitrageurs’ competition to eliminate price manipulation but relies on oracle pricing. \cite{Xavier_Ferreira_2023} study verifiable sequencing rules to prevent MEV and find that while some non-zero miner profits remain, their proposed sequencing rule ensures that user transactions remain unaffected. \cite{canidio2024arbitrageurs} develop a game-theoretic model to differentiate front-running from legitimate trades. Their proposed protocol reduces front-running risk but requires additional message exchanges, which could impact efficiency. \cite{zhou2021a2mm} proposes a modification of AMM algorithms that incorporate optimal on-chain swap routing and arbitrage. In contrast our proposed solution requires only information that already exists in the market. 
    \input{sections/literature}

\section{Our Benchmark: The Constant Product Market Maker (CPMM)}

    \input{sections/CPMM}

\section{The Analysis of Global Market Makers}

    \input{sections/analysis}



\section{Properties of the GMM}

    \input{sections/properties}

\section{Empirical Evaluation}

\input{sections/Data}

\section{Implementation of the GMM}

    \input{sections/Implementation}

\section{Conclusion}

   \input{sections/conclusion}


%
%
%

\newpage
\sloppy
%

\bibliography{aux/refs}

\end{document}

%% file: sections/intro.tex
As cryptocurrentcies' popularity grows, so it does the need to exchange them for other cryptocurrencies and for fiat currencies. Crypto exchanges have emerged to satisfy these needs. A crypto exchange is a platform that allows users to buy, sell, and trade cryptocurrencies. Crypto exchanges can be centralized or decentralized. Centralized EXchanges (CEXs) are owned and operated by a single entity, which holds users' funds and executes trades. Decentralized EXchanges (DEXs) do not rely on a central authority to hold users' funds or execute trades. Instead, they use smart contracts running over a public blockchain such as Ethereum, to automate the trading process. DEXs have become increasingly popular reaching a daily trading volume of billions of dollars \cite{dex-vol}.

There are two main types of DEXs, namely, order book DEXs and Automated Market Makers (AMMs). Order book DEXs were introduced first but nowadays AMMs are the most common type of DEX. Order book DEXs are similar to traditional order book exchanges. Users place orders to buy or sell assets, and these orders are matched with each other to execute trades. AMMs work by creating liquidity pools, which are essentially large baskets of assets that are held by the exchange. When a user wants to trade an asset, they do not actually trade with another user. Instead, they trade with the liquidity pool. The AMM will then use a predetermined mathematical formula to determine the price of the trade. One of the most widely used AMMs is the Constant Product Market Maker (CPMM). CPMMs use a constant product formula, that ensures that the product of the reserves of two assets in the liquidity pool remains constant. The CPMM's formula inputs the volume of the trade and the available liquidity and outputs the price of the trade. 

A key feature of the CPMM algorithm is that the price it computes reflects the relative scarcity of an asset in the AMM's liquidity pool. This effectively translates traders' demand into prices: when market sentiment about an asset's price rises, traders buy the asset from the AMM, reducing its availability in the liquidity pool until the AMM's price aligns with market sentiment.

Market fragmentation across multiple AMMs presents challenges, the most notable being the reliance on costly arbitrage to equalize prices. Each CPMM independently adjusts swap rates based on its own trading history, discovering asset prices at different paces. If two CPMMs start with the same exchange rate and one executes a trade, their rates will diverge, with larger trades causing greater price differences. These discrepancies create arbitrage opportunities, systematically exploited by arbitrageurs to restore price consistency. A recent study \cite{arb-vol} estimates that arbitrage profits on the Ethereum blockchain alone amount to \$qgv$170$ million per year, leading to higher costs for traders and reduced profits for liquidity providers, and thus reducing the efficiency of the blockchain ecosystem.

Another consequence of market fragmentation is increased slippage, i.e. the price's sensitivity to order size. This effect is particularly pronounced in AMMs with small liquidity pools, where large orders significantly alter asset scarcity, causing sharp price impacts in the CPMM. Clearly, aggregating all liquidity pools into a single AMM would substantially reduce slippage if the CPMM algorithm remains in use.

Standard microeconomic analysis suggests that slippage reduces market efficiency as it deviates from the price-taking assumption; see \cite{Aldridge2022}. In practice, however, a more pressing issue is that higher slippage increases the profitability of sandwich attacks. In this front-running strategy, an attacker places a buy order before and a sell order after a victim’s trade, manipulating the price to profit from the slippage they induce. As a result, the victim receives a worse price, while the attacker exploits the price difference for profit.

In this paper, we explore how to address these inefficiencies by computing prices in the spirit of the CPMM but based on the relative scarcity of assets in the aggregate liquidity pools of all AMMs rather than in individual AMM pools. This approach allows prices in a fragmented AMM ecosystem to behave as if they were set by a single AMM.

Our first observation is the failure of the naive approach of applying the CPMM formula using aggregate liquidity pools instead of individual ones, what we call the {\it naive Global Market Maker} (nGMM). This algorithm can be easily manipulated by strategically rearranging transactions, potentially leading to the depletion of the AMM reserves.

Next, we characterize algorithms that fall between the standard CPMM and the nGMM but are not manipulable as described in the previous paragraph. Among these, we define the \textit{Global Market Maker (GMM)} as the algorithm closest to the nGMM.

In the rest of the paper, we study the properties of the GMM and compare it to the CPMM. We show that the GMM eliminates all arbitrage opportunities and that the profits arbitrageurs would earn when all AMMs use the CPMM algorithm instead are shared between the AMMs and the traders. 

We also investigate, both theoretically and numerically, how much the GMM reduces the profitability of sandwich attacks compared to the CPMM. In all simulated examples using real data, we observe reductions of over 50\%.

Finally, we analyze the performance of the GMM with respect to impermanent loss. This is defined, see \cite{Hafner2024}, as the loss incurred by a liquidity provider when depositing assets into an AMM, if the relative prices of those assets change. Since the GMM incorporates liquidity information from other AMMs, the price it offers adjusts before reaching the AMM which minimises impermanent losses.

The rest of the paper is structured as follows. Section \ref{sec:lit} discuss the related literature. Section \ref{secCPMM} introduces the currently most used algorithm, the CPMM, and discuss its shortcomings. Section \ref{sec:GMM} analyses our propososal of improvement of the design of the CPMM, the GMM, and Section \ref{sec:prop} illustrates how the GMM increases performance relative to the GMM. Section \ref{sec:con} concludes.

%% file: sections/literature.tex
\label{sec:lit}

%% file: sections/CPMM.tex
\label{secCPMM}
An automated market maker (AMM) is a smart contract that implements a decentralized exchange protocol based on an algorithm. These protocols allow users, known as \textit{liquidity providers} (LPs), to deposit pairs of assets into liquidity pools, which are then used to execute trades automatically according to the AMM algorithm.

In this section, we describe the algorithm that we use as the starting point of our analysis. This is the most popular algorithm used by AMMs: the constant product market maker (CPMM) introduced by \cite{dex-vitalik}. Hayden Adams provided the first implementation of the CPMM in his decentralized exchange, Uniswap, in 2017, and Uniswap quickly became the most popular decentralized exchange on Ethereum, with CPMMs emerging as the dominant type of AMM.

The Constant Product Market Maker (CPMM) manages a pool of reserves consisting of two assets. A trader who wishes to swap one asset for another sends a certain amount of the first asset into the AMM. The CPMM algorithm then calculates the amount of the second asset to be returned to the trader, ensuring that the product of the reserves of both assets in the liquidity pool remains constant.

Formally, let $x_i$ and $y_i$ be the reserves of assets $X$ and $Y$ in the algorithm's liquidity pool. If a trader sends an amount $\Delta x_i>0$ of asset\footnote{Since $X$ and $Y$ are arbitrary assets, it is sufficient to describe orders in which the trader sends asset $X$. The case in which the trader sends asset $Y$ can be handled by relabeling the assets.} $X$ into the algorithm, the CPMM returns an amount $\Delta y_i>0$ of asset $Y$ such that:
\begin{equation}
x_i\cdot y_i=(x_i+\Delta x_i)\cdot (y_i-\Delta y_i).
\label{eq:CPMM}    
\end{equation}
Solving this equation for $\Delta y_i$ allows for a characterization in terms of the amount of $Y$ obtained by a trader that sends $\Delta x_i$ units of $X$ to the CPMM:
\begin{equation}
\Delta y_{CPMM}(\Delta x_i;x_i,y_i) = \frac{y_i} {x_i+\Delta x_i} \Delta x_i=\frac{r_i}{1+\frac{\Delta x_i}{x_i}}\Delta x_i,
\label{eq:CPMMbis}
\end{equation}
where $r_i\equiv \frac{y_i}{x_i}$ is the ratio of reserves of the AMM. The interpretation of the last equation is that the terms of trade $\frac{\Delta y}{\Delta x}$ of the CPMM are given by a measure of the relative scarcity of the assets in the liquidity pool, adjusted by a measure of the order's size relative to the liquidity pool of the AMM. The first term represents the {\it marginal price}, as it corresponds to the limit of the terms of trade when $\Delta x_i$ tends to zero. The second term captures {\it slippage}, which quantifies the rate at which the terms of trade deteriorate as the order size increases relative to the liquidity pool.

We identify next some drawbacks of the CPMM when it operates in a fragmented ecosystem with several AMMs, all using the CPMM algorithm. The first and most obvious is that differences in prices arise naturally if traders do not split the orders optimally between the different AMMs. This creates profitable arbitrage opportunities that become a cost for traders as we illustrate in the next example.

\begin{examplebox}{Toy example - Part 1: Arbitrage between CPMMs.}
    Consider two AMMs that use the CPMM algorithm, each with a liquidity pool containing 100 ETH and 400,000 UST. The marginal price is the same for both CPMMs and equal to 4,000 UST/ETH. All the computations in this example are done using \eqref{eq:CPMMbis}.
    
    Suppose that a trader requires 10ETH. He can get them from AMM 1 sending 44,444 UST to AMM 1. After the trade, AMM 1's pool will contain 90 ETH and 444,444 UST. The marginal price in AMM 1 is now 4,938 UST/ETH. An arbitrage opportunity has emerged.

    An arbitrageur can profit by sending 5 ETH to AMM 1 to get 23392 UST and then sending 21,052 UST to AMM 2 to obtain 5 ETH back. After the arbitrage, the reserves of both CPMMs will be equal to 95 ETH plus 421,052 UST and the arbitrageur makes a profit of 2,339 UST.

    Suppose that after this trade, the trader wants to convert the 10 ETH back into UST. He can send them to AMM 1 to obtain 40,100 UST. Thus the AMM 1 pool will contain 105 ETH and 380,952 UST. Again, an arbitrage operation has emerged which is seized by an arbitrageur who sends 5 ETH to AMM 2 and then sends the UST received to AMM 2. These two trades give the arbitrageur a profit of 2,005 UST. After these two arbitrage operations, the reserves of both AMMs will be restored to 100 ETH and 400,000 UST each.

    Overall, in these 6 operations, the CPMMs retained their initial reserves intact, the arbitrageur has made a profit of 4,344 UST which matches to the net cost to the trader.  
\end{examplebox}

Thus, whereas arbitrage provides the service of equalising the marginal prices across AMMs, it is at a cost for the trader(s) which is equal to the profits of the arbitrageurs. 

The second implication of the fragmentation of the reserves into different AMMs is that each individual AMM only has a portion of total liquidity reserves which makes slippage substantially larger than in a hypothetical AMM holding the aggregate reserves of all AMMs. Higher slippage makes it more profitable to conduct a {\it sandwich attack}. This is a manipulative attack that combines two trades, one before the target transaction (frontrunning) and another one after (backrunning) to profit from the price movement caused by the victim’s trade. In CPMMs, this involves inserting trades around a large swap to capitalize on slippage, forcing the victim to execute their trade at a worse price while the attacker profits from the price difference. We illustrate sandwich attacks in the next example.

 \begin{examplebox}{Toy example - Part 2: MEV extraction in CPMM.}
    Consider an AMM using the CPMM algoritm and with a liquidity pool containing 100 ETH and 400,000 UST. The marginal price of ETH is then 4,000 UST/ETH. Suppose that a trader interested in converting 40,000 UST in ETH sends them to the AMM. The MEV-extractor can perform a sandwich attack by sending 60,000 UST just before the victim's transaction is executed to send back immediately after the ETH obtained. This would work as follows:

\begin{itemize}
    
\item \textbf{Frontrunning:} MEV-extractor transaction 1 sends 60,000 UST to get 13.0435 ETH in return, and changes the vector of reserves of AMM 1 to 86,9565 ETH and 460.000 UST. The marginal price of ETH increases to 5,290 UST/ETH.

\item \textbf{Victim's transaction:} The victim sends 40,000 UST to obtain in return 6.9565 ETH. The resulting reserves of the AMM are then 80 ETH and 500,000 UST. Note that in the absence of the frontrunning transaction, the victim would have received 9,0909 ETH, or equivalently, they only needs to send 29,906 UST to get  6,9565 ETH, i.e. they pay an overpriced of 10,094 UST.

\item \textbf{Backrunning:} MEV-extractor transaction 2 sends 13.0435 ETH to get 70.094 UST. The AMM's reserves after the trade are 110 ETH and 363,636 UST. The MEV-extractor obtains a total profit of 10,094 UST, which is equal to the trader’s overprice.

\end{itemize} 
\end{examplebox}

A simple analytical exercise allow us to quantify the profits derived by a general MEV sandwich operation consisting of submitting a frontrunning order $\hat \Delta x$ to a traders order $\Delta x$ followed by a backrunning order in which the manipulator sends back to the AMM the amount obtained in the frontrunning trade. We refer to this manipulative operation as an {\it elementary MEV sandwich $\hat \Delta x$ to order $\Delta x$}. The next result provides the corresponding analytical formula for the profits.

\begin{proposition}
    The profit of an elementary MEV sandwich $\hat \Delta x$ to a trader order $\Delta x$ to an AMM i with liquidity reserves $(x_i,y_i)$ and running the CPMM algorithm is equal to:
    \begin{equation}
        \left( \frac{ \left(1+ \frac{\hat \Delta x}{x_i}+\frac{ \Delta x}{x_i} \right)^2 }{  \left(1+ \frac{\hat \Delta x}{x_i}+\frac{ \Delta x}{x_i} \right)\left(1+ \frac{\hat \Delta x}{x_i} \right)-\frac{\Delta x}{x_i}} -1 \right) \hat \Delta x
    \end{equation}
\end{proposition}

\begin{proof}
    Since the proof of this proposition is a particular case of Proposition \label{proMEVGMM} for the case $x_i=x$ we postpone the proof.
\end{proof}

\begin{figure}[ht!]
    \centering
    \includegraphics[width=10cm]{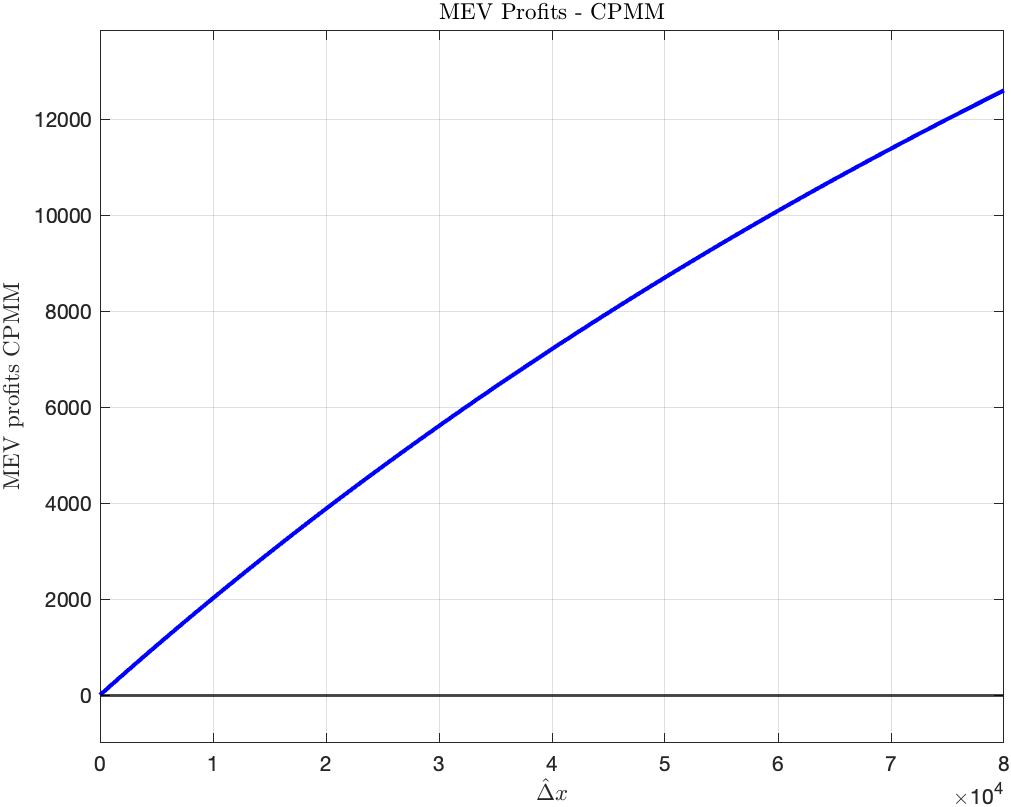} 
    \caption{The graph plots the profits of elementary MEV sandwich manipulations $\hat \Delta x$ for $x_i=400,000$ and $\Delta x =40,000$.}
    \label{fig:MEV_CPMM}
\end{figure}

Figure \ref{fig:MEV_CPMM} illustrates the proposition for different values of $\hat \Delta x$.

Finally, the fragmentation of the AMM ecosystem, together with the CPMM algorithm, means that AMMs often operate with outdated prices, leading to larger \textit{impermanent losses}. This is the loss liquidity providers may experience when supplying liquidity to AMMs due to the automatic rebalancing of token holdings in the liquidity pool as the relative price of tokens fluctuates. Specifically, as one token’s price decreases relative to the other, liquidity providers end up holding more of the depreciated asset and less of the appreciated one than they initially deposited. The term \textit{impermanent} reflects that the loss may be reversed if the price ratio returns to its original state. However, in practice, such reversals are rare, making impermanent loss a crucial factor for liquidity providers to consider when engaging with AMMs.

\begin{figure}[!ht]
    \centering
    \includegraphics[width=10cm]{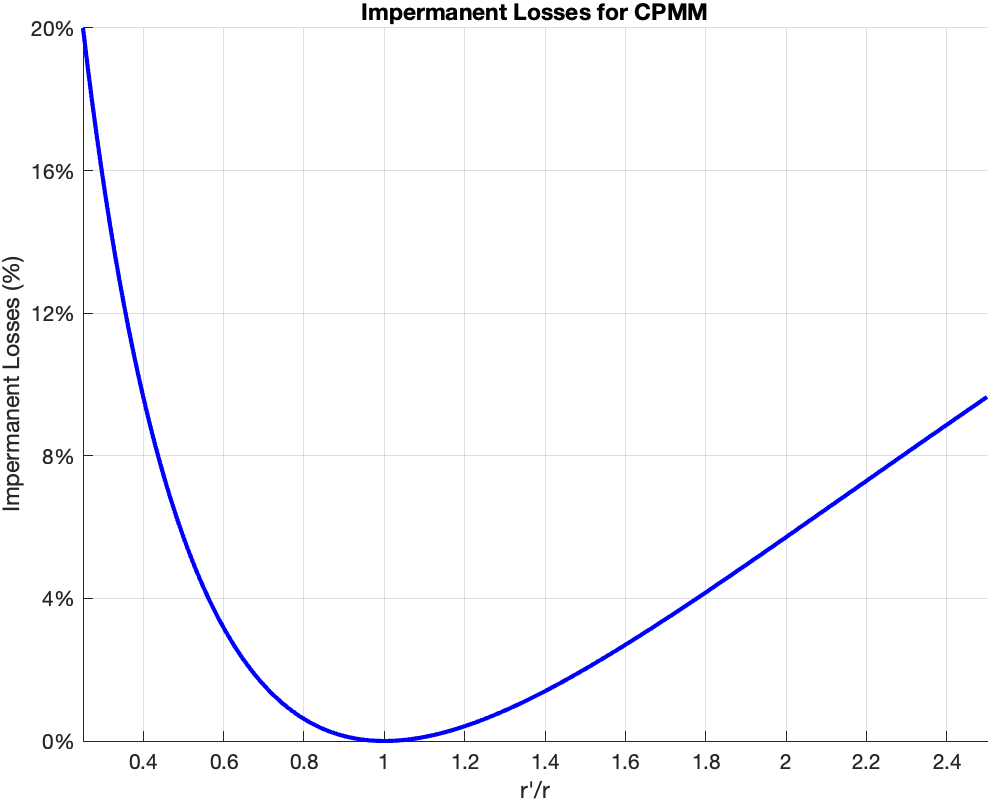} 
    \caption{The graph plots the impermanent losses as a function of the price increase $\frac{r}{r'}$.}
    \label{fig:GMM_IL_CPMM}
\end{figure}

Mathematically, impermanent loss is defined as the relative loss incurred by an LP compared to simply holding the original asset allocation:\footnote{See  \cite{lipton2024unified} for a comprehensive discussion of alternative measures of impermanent losses. Ours correspond to what they call relative IL of borrowed Liquidity Provision.}
\[
\text{IL}_i = 1 - \frac{V_{i}}{V^0_{i}}
\]
where \( V_{i,t}\equiv y_i+r x_i \) is the value of the liquidity holdings of the AMM at the current price, and \( V^0_i\equiv y_i^0+r x_i^0 \) is the value of the liquidity holdings if the liquidity provider had kept the initial asset allocation. The impermanent losses for the CPMM has a simple expression under the assumption that the initial ratio of reserves of the AMM and the final ratio of reserves are equal to $r'$ and $r$, respectively, and so are the initial and final marginal prices of the AMM:
\begin{equation}
\text{IL}_i = 1 - \frac{2}{\sqrt{\frac{r}{r'}} + \sqrt{\frac{r'}{r}}}
\label{eq:ILCPMM}
\end{equation}
where \( r' \) and \( r \) are the initial and final price ratios of the token pair, respectively. Figure \ref{fig:GMM_IL_CPMM} provides a graph of the impermament losses as a function of the ratio $\frac{r}{r'}$ and the next example illustrates the concept.

\begin{examplebox}{Toy example - Part 3: Impermanent loss in CPMM.}
     Consider two automated market makers AMM 1 and AMM 2 using CPMM algorithm, each with a liquidity pool with 100 ETH and 400000 UST. The marginal price is then 4,000 UST/ETH. 
     
     Suppose that an event makes traders in the market believe that the new price should be 3,000 UST/ETH. As a result, they sell ETH to each of the two AMMs until the marginal price moves to 3,000 UST/ETH in both CPMMs. The new reserves for each CPMM after the transactions are $115.47$ ETH and  $346,410.16$  UST.

We compute the impermanent loss of each AMM as the difference between the value of the CPMM’s original reserves valued at the new price ($V_i^0$) and the current reserves valued at the new price ($V_i$) normalised by the former value, i.e.,
\begin{eqnarray*}
V_i^0 &= & 100 \text{ ETH} \times 3,000 \text{ UST/ETH} + 400,000 \text{ UST}, \\
V_i &=& 115.47 \text{ ETH} \times 3,000 \text{ UST/ETH} + 346,410.16 \text{ UST}, \\
IL_i &=& \frac{V_i^0 - V_i}{V_i^0} = \frac{700,000 UST-692,820.16 UST}{700,000 UST}=1.03\%.
\end{eqnarray*}
    \end{examplebox}

%% file: sections/analysis.tex
\label{sec:GMM}


In this section, we begin by observing that the issues of the CPMM, as identified in the previous section, could be mitigated if the AMM made better use of the information available on blockchains like Ethereum, particularly the liquidity pools of other AMMs. Arbitrage could be eliminated by computing the AMM price in relation to existing market prices, which, under the CPMM design, can be determined simply by checking reserve balances. The profitability of sandwich attacks depends on slippage, which, in turn, decreases as the AMM’s reserve size increases—at least in the case of the CPMM. If prices were instead computed using aggregate reserves, they would become less sensitive to large orders, thereby reducing slippage and, consequently, the profitability of sandwich attacks. Finally, if prices adjust before a trade order reaches the AMM, they can respond more quickly to market conditions, helping to minimize impermanent losses.


In our formalization, we assume a set $I$ of AMMs, each with liquidity pools described by the vector $(\vec{x},\vec{y})\in \real_+^{2I}$, where
\[
x \equiv \sum_{j\in I} x_j, \quad \text{and} \quad x_{-i} \equiv \sum_{j\in I\setminus\{i\}} x_j,
\]
with analogous definitions for $y$ and $y_{-i}$. Let also $r=\frac{y}{x}$ be the global ratio of reserves.

We are interested in algorithms that determine the terms of trade using the information contained in the vector of liquidity reserves or all AMMs. We refer to such algorithms as {\it global}. An example is the global algorithm that mimics the behavior of a CPMM algorithm with the aggregate reserves.

\begin{definition}
Given a vector of liquidity pools $(x,y)$, the {\it naive Global Market Maker} (nGMM) algorithm swaps $\Delta x_i>0$ of $X$ for an amount of $Y$:
\begin{equation}
\Delta y_{\text{nGMM}}(\Delta x_i;x,y) = \frac{y}{x+\Delta x_i} \Delta x_i=\frac{r}{1+\frac{\Delta x_i}{x}} \Delta x_i,
\label{eq:nGPMM}
\end{equation}
if less than $y_i$, and $\Delta y_{\text{nGMM}}(\Delta x_i;x,y)=y_i$, otherwise.
\end{definition}

This algorithm determines the terms of trade as if executing a hypothetical CPMM with reserves equal to the sum of individual reserves. Consequently, the terms of trade reflect both global scarcity and global slippage. Notably, if all AMMs adopt this algorithm, arbitrage opportunities are eliminated, as illustrated in the following example, and transformed into better terms of trade for traders.

 \begin{examplebox}{Toy example - Part 4: nGMM}
     Consider two automated market makers AMM 1 and AMM 2 using CPMM algorithm, AMM 1's pool contains 90 ETH and 444,444 UST and AMM 2's pool 100 ETH and 400000 UST. This is exactly the distribution of reserves in Part 1 before arbitrage is executed.
     
     However, no strictly profitable arbitrage opportunities exist here: if an arbitrageur sends 5 ETH to AMM 2 gets 21,652 UST that after sending them to AMM 1 delivers 5 ETH. Thus, zero net trade.
     
     Suppose instead that the a trader sends 10 ETH to AMM 1 when AMM 1's pool contains 90 ETH and 444,444 UST and AMM 2's pool 100 ETH and 400000 UST. The trader gets 42,222 UST in return. This is 2,122 ETH more than in the CPMM of Part 1 after arbitrage.
\end{examplebox}
However, the nGMM has a significant drawback:\footnote{Another drawback of the nGMM, which we do not discuss extensively, is that it can lead to the complete depletion of one asset’s reserves, forcing the AMM to cease operations. Neither the CPMM nor the algorithm we propose below, the GMM, share this issue.} it can be exploited by an strategic party that submits a specific sequences of transactions in the market. This issue is illustrated in the following example.

\begin{examplebox}{Toy example - Part 5: Exploiting nGMM}
Consider two automated market makers AMM 1 and AMM 2 using nGMM, each with a liquidity pool with 100 ETH and 400000 UST. Consider the effect of the following transactions:

     \begin{itemize}
     \item {\bf Transaction 1:} Send 10 ETH to AMM 1 and gets 38,095 UST in return. The resulting reserves of AMM 1 are 110 ETH and 361,905 UST. AMM 2 still has 100 ETH and 400,000 UST.  

     \item {\bf Transaction 2:} Send 10 ETH to AMM 2 and gets 34,632 UST in return. The resulting reserves of AMM 2 are then 110 ETH and 365,368 UST. AMM 1 are 110 ETH and 361,905 UST.

     \item {\bf Transaction 3:} Send 38,095 UST to AMM 1 to get 10.95 ETH. The reserves for AMM 1 would then be 99,05 ETH and 400,000 UST, which is strictly less than its initial reserves (same amount of UST and less of ETH). AMM 2 remains with reserves 110 ETH and 365,368 UST.

     \item {\bf Transaction 4:} Send 34,632 UST to AMM 2 to get 9.05 ETH in return. The reserves for AMM 2 would then be 100.95 ETH and 400,000 UST, which is strictly greater than its initial reserves (same amount of UST and greater of ETH). 

     \end{itemize}
AMM 2 gains 0.95 ETH that have been obtained from AMM 1. The net variation of trades A, B, C and D is equal to zero.
\end{examplebox}

In this example, AMM 1 suffers a net decline in its holdings because the nGMM sells UST at a low price when it is relatively abundant among the two AMMs, only to repurchase it later at a higher price since

the sale of UST by AMM 2 has made it relatively more scarce. This net decline in AMM 1 holdings correspond to an equal increase in the holdings of AMM 2 because the orders have zero net flow. More generally, exploitation may take a more sophisticated form and involve other agents. The key feature is that the AMM using the nGMM algorithm suffers a net decline in its holdings after a particular sequence of transactions. Since the decline of the AMM is a gain for the other agents affected by the sequence of transactions, at least one of them can do strictly better by submitting this sequence of transactions because their gain must be which unavoidably results into an identical increase in the net position of all the other agents affected by the transactions

Consequently, one only wants to consider global algorithms for which no sequence of trades lead to a net decline in its holdings. To formalise this condition, 

This condition captures a minimal requirement for non-exploitability, asserting that no sequence of orders should be able to move AMM \( i \) from its original reserve position to a strictly worse one through a cycle of interactions. It is important to note that the definition is intentionally agnostic about the mechanisms or strategies employed by other AMMs or agents in the system. While some algorithms might not even allow a transition from \( (\vec{x}^0, \vec{y}^0) \) to \( (\vec{x}, \vec{y}) \), the definition requires only the existence of such a transition, which simplifies the analysis by avoiding the need to track intermediate transactions or specify the behavior of other market participants.

\begin{definition}
A global algorithm employed by an AMM \( i \) is said to be \emph{exploitable} if there exist an initial distribution of liquidity pools \( (\vec{x}^0, \vec{y}^0) \) and a final distribution \( (\vec{x}, \vec{y}) \) such that:
\begin{itemize}
    \item There exists an order sent to AMM \( i \) that, starting from the initial distribution, moves AMM \( i \)'s reserves to \( (x_i, y_i) \).
    \item There exists another order sent to AMM \( i \) that, starting from the final distribution, moves AMM \( i \)'s reserves to\footnote{We denote \( (x_i, y_i) \leq (x_i', y_i') \) if \( x_i \leq x_i' \) and \( y_i \leq y_i' \), with at least one inequality being strict.} \( (x_i', y_i') \leq (x^0_i, y^0_i) \).
\end{itemize}
\end{definition}

This condition captures a minimal requirement for non-exploitability, asserting that no sequence of orders should be able to move AMM \( i \) from its original reserve position to a strictly worse one through a cycle of interactions. It is important to note that the definition is intentionally agnostic about the mechanisms or strategies employed by other AMMs or agents in the system. While some algorithms might not even allow a transition from \( (\vec{x}^0, \vec{y}^0) \) to \( (\vec{x}, \vec{y}) \), the definition requires only the existence of such a transition, which simplifies the analysis by avoiding the need to track intermediate transactions or specify the behavior of other market participants. 

Moreover, the definition is deliberately neutral with respect to the identity of the strategic actor: it does not matter who executes the trades, only that AMM \( i \) incurs a loss. In a closed system, any loss in reserves by one participant necessarily implies that the reserves have been absorbed elsewhere. Therefore, the mere existence of a loss implies the presence of a trader or another AMM that benefits—either directly or through redistribution—from the transaction sequence. This justifies the notion of exploitability and underscores the importance of designing AMM algorithms that are robust against such reserve-depleting sequences.

An obvious approach to achieving non-exploitability is to systematically reduce the amount of assets returned to traders. However, this approach is unappealing, as it assumes that doing so incurs no cost to the AMM. In reality, worsening the terms of trade for traders is likely to reduce their participation. Therefore, it makes sense to require non-exploitability while also ensuring that the terms of trade for traders do not deteriorate. Indeed, this aligns with the spirit of our objective: improving the efficiency of the blockchain ecosystem so that both AMMs and traders benefit.  

To address this question we start with a few preliminary observations. First, we adopt the point of view that both AMMs and traders improve absorbing part of the arbitrageurs profits. Thus, the benchmark of comparison must have arbitrage. Second, an algorithm that makes arbitrage less profitable must worsen terms of trade. If successful in eliminating arbitrage, this worsening of the terms of trade should affect traders and thus hinder the possibility of improving traders welfare. Except this worsening of the terms of trade in one AMM is compensated with betters terms of trade in another AMM, which suggests that achiving an improvement for trades requires modeling how they route their orders across AMMs. Here we shall assume that traders always trade with the AMM offering the best terms of trade.

\begin{definition}
    A global algorithm is trade preserving with respect to the CPMM if for any given initial distribution of liquidity $(\vec{x},\vec{y})$, a trader interested in swapping $\Delta x$ units of $X$ after all arbitrage opportunities have been exhausted can obtain better terms of trade when all the AMMs use the global algorithm than when they all use the CPMM algorithm.   
\end{definition}

At this level of generality, it is necessary to impose some minimum properties on the global algorithms to deduce meaningful characterizations. 

\begin{definition}
    A global algorithm is natural if:
    \begin{itemize}
        \item[(i)] It offers better terms of trade to traders when its reserve increase while the ratio of reserves and the other AMMs reserves remain constant.
        \item[(ii)] It does not display strictly profitable arbitrage opportunities when all AMMs have the same ratio of reserves.
    \end{itemize}
\end{definition}

The first property simply says that slipagge decreases with the size of the AMM and the second that arbitrage opportunites can only arise from differences in the ratio of reserves. Both are natural properties that generalize the CPMM.

\begin{proposition}
    A natural global algorithm is trade preserving and not exploitable only if for any possible distribution of reserves and swap it does not return the trader more assets than the CPMM.
\end{proposition}

\begin{proof}
    By contradiction, suppose a distribution of reserves $(\vec{x}^0,\vec{y}^0)$ and a swap that $\Delta x_i>0$ of $X$ that gets $\Delta y_i$ units of $Y$ in return and that return the trader more assets than the CPMM. The definition of CPMM means that the swap $\Delta x_i$ decreases the product of reserves, this is,
    \begin{equation}
    x_i^0y_i^0>x_iy_i,    
    \label{eqxy0}
    \end{equation}
    for $x_i\equiv x_i^0+\Delta x_i$ and $y_i\equiv y_i^0-\Delta y_i$. Now suppose a distribution of reserves $(\vec{x},\vec{y})$ such that the other AMMs have the same ratio of reserves as $i$, i.e. $\frac{y_i}{x_i}=\frac{y_j}{x_j}$, and that $i$ has the largest product of reserves, i.e. $x_iy_i=\max_j x_jy_j$. The first property means that no arbitrage opportunities are exploited before the trader trades and the second that AMM $i$ is the most convenient AMM for the trader. Thus, trade preserving means that the terms of trade for the trader cannot be worse than the CPMM, which together with the definition of CPMM means that after any trade the product of reserves of AMM $i$ cannot increase. Thus if the trader submits $\Delta x$, AMM $i$ ends up with $x^0_i$ units of $X$ and a number of units of $Y$ strictly lower than $y_0$, which contradicts that the algorithm is not exploitable.
\end{proof}

Consequently, we propose the following algorithm. 

\begin{definition}
    The Global Market Maker (GMM) algorithm is defined by:
    \begin{equation}
\Delta y_{GMM}(\Delta x;x,y) = \min\{ \Delta y_{CPMM}(\Delta x;x_1,y_1),\Delta y_{nGMM}(\Delta x;x,y) \}   
\label{eq:GPMM}
\end{equation}
\end{definition}

\begin{figure}[ht!]
    \centering
    \includegraphics[width=12cm]{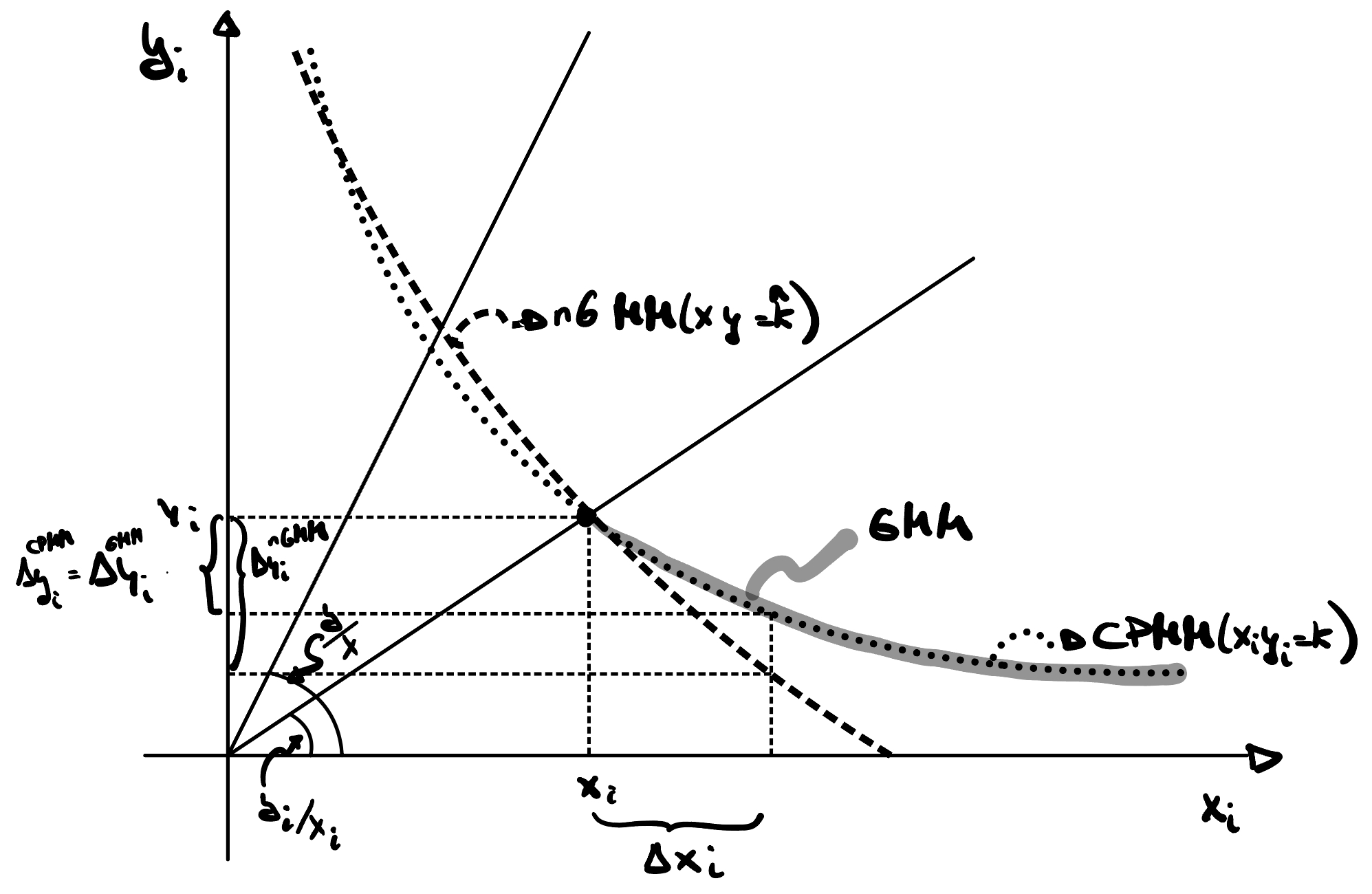} 
    \caption{The graph illustrates a divergent swap where the GMM receives $\Delta x_i$ units of $X$ and returns $\Delta y_i^{GMM}$ of $Y$ to the trader, compared to the returns in the CPMM ($\Delta y_i^{CPMM}$) and nGMM ($\Delta y_i^{nGMM}$) cases. It also shows the possible liquidity reserves of AMM $i$ after a swap, depending on the algorithm used (CPMM, nGMM, GMMM).}
    \label{fig:GMM_div}
\end{figure}

To understand how the GMM determines terms of trade, it is useful to distinguish whether a swap $\Delta x_i>0$ moves the AMM’s reserve ratio away from that of the other AMMs, which we call a {\it divergent swap}. This occurs when the initial vector of reserves satisfies $\frac{y_i}{x_i} \leq \frac{y_{-i}}{x_{-i}}$. In this case, asset $X$ is relatively less scarce locally than globally, causing the nGMM to pay a higher price $\Delta y^{nGMM}_i$ for $X$ than the CPMM price $\Delta y^{CPMM}_i$. Consequently, the GMM price $\Delta y^{GMM}_i$ coincides with the CPMM price. This is illustrated in Figure \ref{fig:GMM_div}.

\begin{figure}[ht!]
    \centering
    \includegraphics[width=12cm]{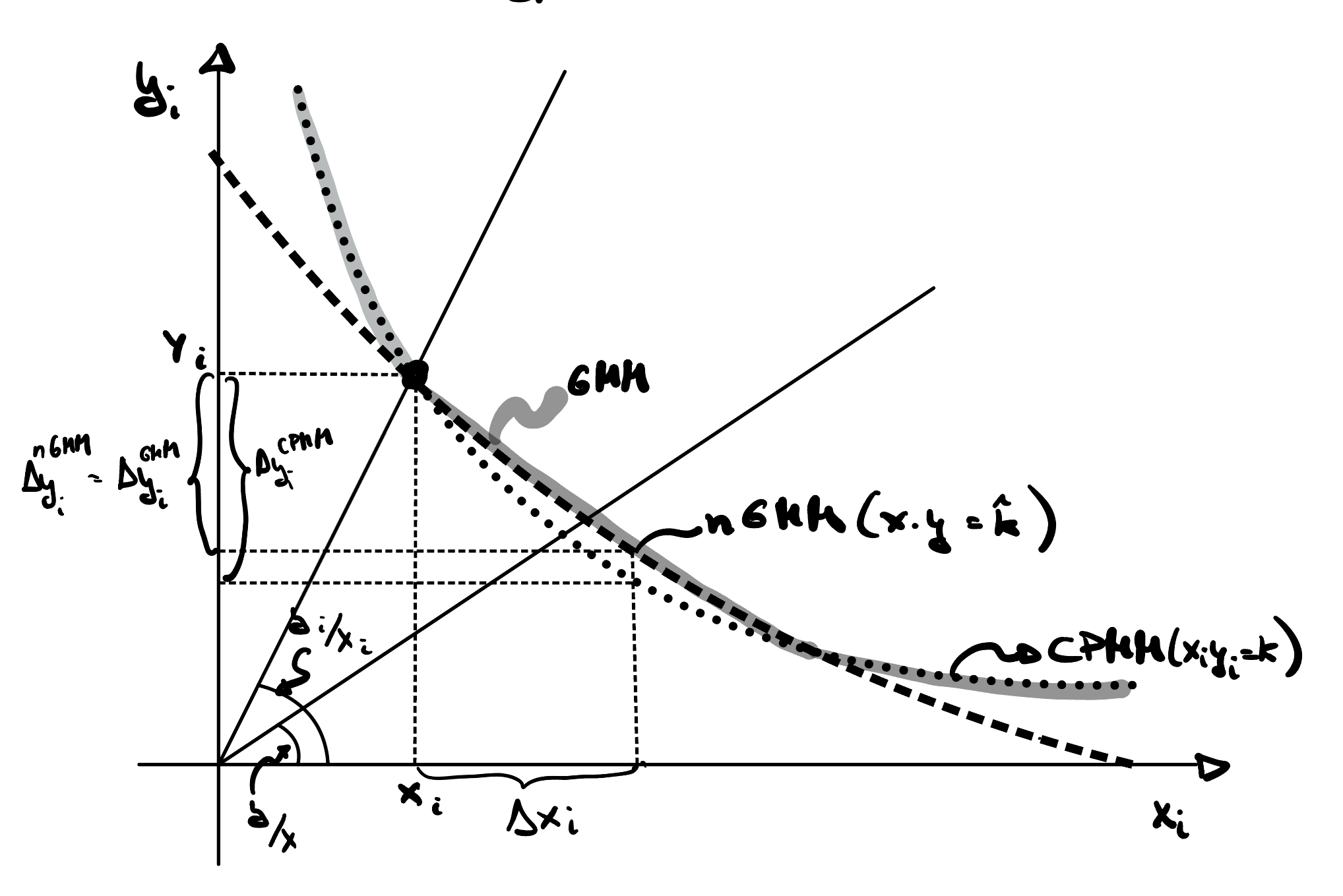} 
    \caption{The graph illustrates a convergent swap where the GMM receives $\Delta x_i$ units of $X$ and returns $\Delta y_i^{GMM}$ of $Y$ to the trader, compared to the returns in the CPMM ($\Delta y_i^{CPMM}$) and nGMM ($\Delta y_i^{nGMM}$) cases. It also shows the possible liquidity reserves of AMM $i$ after a swap, depending on the algorithm used (CPMM, nGMM, GMMM).}
    \label{fig:GMM_conv} 
\end{figure}

A swap $\Delta x_i>0$ is {\it non-divergent} when $\frac{y_i}{x_i} > \frac{y_{-i}}{x_{-i}}$. There are two possible cases. The first, which we refer to (with a slight abuse of terminology) as a {\it convergent swap}, occurs when the nGMM returns less $Y$ than the CPMM, causing the GMM to coincide with the nGMM. This happens when $\Delta x_i$ is small enough that the swap moves the reserve ratio toward that of the other AMMs, justifying the name. However, it can also occur if $\Delta x_i$ is large enough to decrease the reserve ratio below that of the other AMMs but not so large that the nGMM price exceeds the CPMM price. This is illustrated in Figure \ref{fig:GMM_conv}.

\begin{figure}[ht!]
    \centering
    \includegraphics[width=12cm]{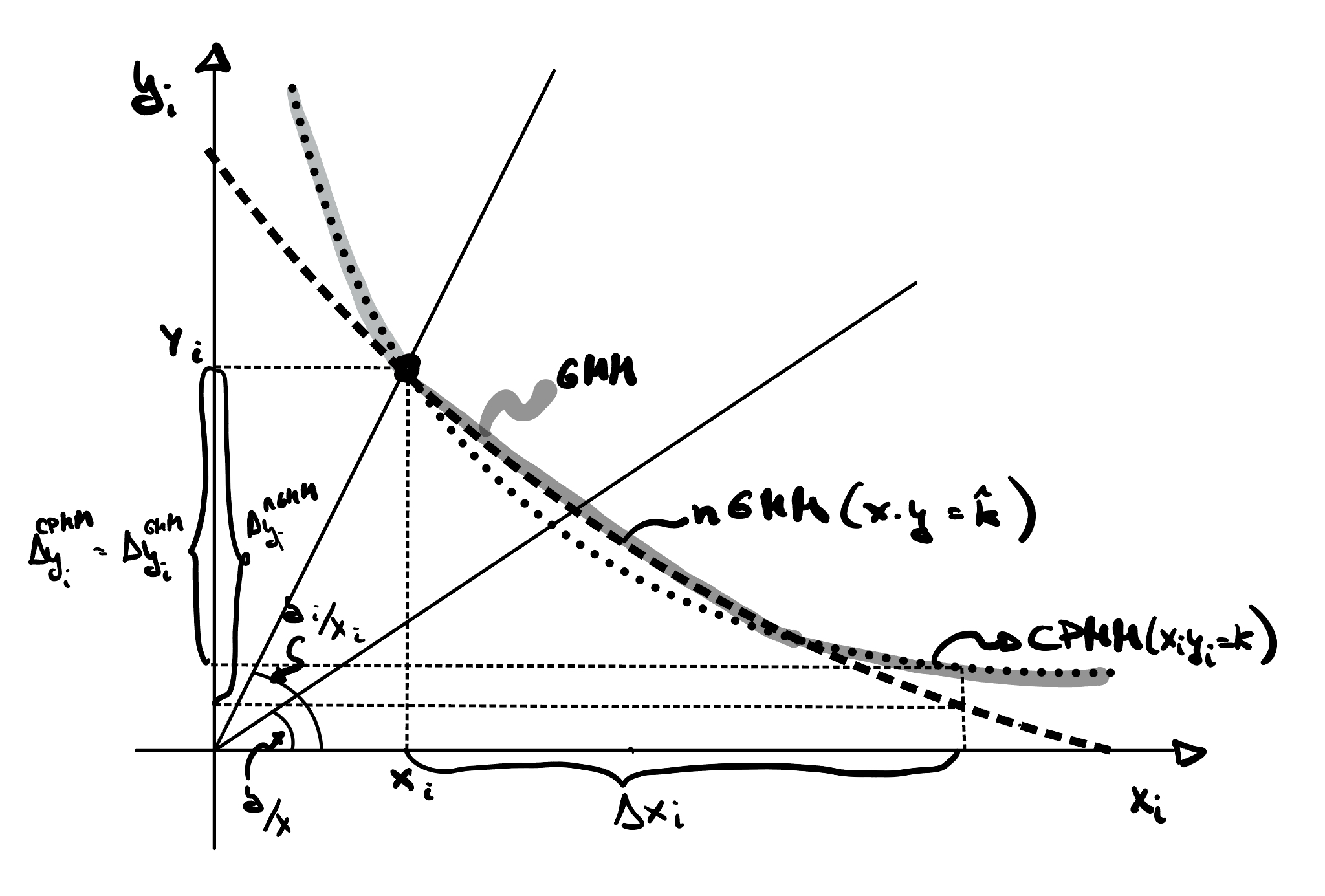} 
    \caption{The graph illustrates an overshooting swap where the GMM receives $\Delta x_i$ units of $X$ and returns $\Delta y_i^{GMM}$ of $Y$ to the trader, compared to the returns in the CPMM ($\Delta y_i^{CPMM}$) and nGMM ($\Delta y_i^{nGMM}$) cases. It also shows the possible liquidity reserves of AMM $i$ after a swap, depending on the algorithm used (CPMM, nGMM, GMMM).}
    \label{fig:GMM_over} 
\end{figure}

The last case, {\it the overshooting swap}, is when the non divergent swap $\Delta x$ is so large that the nGMM pays a higher price $\Delta y^{nGMM}_i$ for $X$ than the CPMM price $\Delta y^{CPMM}_i$, causing the GMM price $\Delta y^{GMM}_i$ to match the CPMM price.\footnote{Formally, the overshooting case occurs at the point where:
\[
\frac{y_i}{x_i} \cdot \frac{y_i-\Delta y_{CPMM}(\Delta x_i;x_i,y_i)}{x_i+\Delta x_i} = \frac{y}{x} \cdot \frac{y-\Delta y_{CPMM}(\Delta x_i;x_i,y_i)}{x+\Delta x_i}.
\]
} We refer to this case as an {\it overshooting swap} and illustrate it in Figure \ref{fig:GMM_over}.

Finally, note that the GMM always receives prices that are not worse than the CPMM, strictly better in the case of convergent swaps. This means that the CPMM product of reserves is expected to grow since the CPMM algorithm keeps the product constant. Next corollary formalises this idea.

\begin{corollary}
    The product of reserves of an AMM with the GMM algorithm weakly increases with every swap, strictly if the swap is convergent.
    \label{cor:incprod}
\end{corollary}

That the GMM is natural and not exploitable is direct. We check whether it is trade preserving once we study its properties and in particular whether it displays arbitrage opportunities.

%% file: sections/properties.tex
\label{sec:prop}

In this section, we show that by exploiting efficiently the aggregation of the information from the liquidity pools of the other AMMs, the GMM improves in the following key dimensions of the design of an AMM: first, it eliminates arbitrage opportunities and splits the efficiency gains between traders and AMMs; second, it reduces slippage and thus the profitability of MEV sandwich trades, and third, it reduces impermanent losses. A second order effect is that the share of arbitrage profits that end in the AMMs increase their liquidity pools and as a consequence further reduce their slippage.

\subsection{Arbitrage}

\label{subsec:arb}

We revisit Example 1 to show that in the GMM no arbitrage opportunities appear, and that the profit that arbitrageurs obtain in the CPMM is split between the AMMs and the traders.

\begin{examplebox}{Toy example - Part 6: GMM} 
    
   Consider two AMMs that use the GMM algorithm, each with a liquidity pool containing 100 ETH and 400,000 UST.
 
A trader that sends 44,444 UST to AMM 1 gets 10 ETH in return, which is the minimum between the CPMM return of 10 ETH, and the nGMM return of 10.53 ETH. This is also consistent with the fact that since both GMMs have the same reserves, the trade is divergent. After this trade, as in part 1, GMM1’s pool updates to 90 ETH and 444,444 UST and GMM2's pool remains unchanged. 

No arbitrage opportunities exist because arbitrage requires convergent trades which are priced using the nGMM algorithm. For instance, an arbitrageur that sends 5 ETH to AMM 1, gets the minimum between the CPMM return of 23,392 UST and the nGMM return of 21652 UST. This trade updates AMM 1 pool of reserves to 95 ETH and 422,792 UST. If after that the arbitrageur sends the 21652 UST obtained to AMM 2 gets the minimum of the CPMM return of 5.13 ETH and the nGMM return of 5 ETH. Thus, the arbitrageur's net trade is zero, and no profit is made.

    Suppose now that, similarly to Example 1, the reverse transaction is executed, namely that a trader sends 10 ETH to AMM 1. The trade is convergent and the AMM returns to the trader 42,222 UST and the resulting reserves of GMM2 after the trade are 100 ETH and 402,222 UST.

    After this trade, the reserves of the GMM have increased in 2,222 UST and the trader paid 2,222 UST less than in the CPMM with arbitrage described in Part 1. In other words, GMM splits evenly the profit of the arbitrageur between the trader and the GMM.
    
\end{examplebox}

The insights of the above example hold true in general. 

\begin{proposition}
There are no strictly profitable arbitrage opportunities between a set of AMMs that use the GMM.  
\label{cor:arbitrage}
\end{proposition}

\begin{proof}
Arbitrage consists of a sequence of swaps initiated by the arbitrageur across different AMMs. Regardless of which AMM executes each swap, the GMM always offers worse terms of trade to the arbitrageur than the nGMM. Moreover, the nGMM preserves the product of aggregate reserves. Consequently, we have:
\begin{eqnarray*}
    xy & \leq  & (x+\Delta x_1)(y-\Delta y_1)\\
    & \leq  & (x+\Delta x_1+\Delta x_2)(y-\Delta y_1-\Delta y_2)\\
    & \leq & \dots \\
    & \leq & \left(x+\sum_{t=1}^T\Delta x_t\right)\left(y-\sum_{t=1}^T \Delta y_t\right),       
\end{eqnarray*}
for any sequence of trades sent by the arbitrageur to the different AMMs, where $\Delta x_i$ represents the amount of $X$ moved from the arbitrageur to an AMM, and $\Delta y_i$ represents the amount of $Y$ received by the arbitrageur from the AMM. 

This inequality implies that either $\sum_{t=1}^T \Delta y_t \leq 0$ or $\sum_{t=1}^T\Delta x_t\geq 0$, meaning that the arbitrageur's net position after the swaps,  
\[
\left(-\sum_{t=1}^T\Delta x_t, \sum_{t=1}^T \Delta y_t\right),
\]
must be either zero or have a strictly negative component. Thus, there are no strictly profitable arbitrage opportunities.
\end{proof}

Thus, replacing the CPMM with the GMM could benefit both traders and AMMs by converting strictly profitable arbitrage opportunities into additional profits for them. For any sequence of swaps submitted by traders or arbitrageurs to different AMMs, the total assets transferred to and withdrawn from the AMMs must equal the total variation in liquidity reserves. Arbitrage profits create a wedge between the net flows of traders and AMMs. Eliminating arbitrage removes this wedge, allowing for higher profits for both.

The question that remains is whether the distribution of these profits is sufficient to make the GMM trade preserving.\footnote{An additional benefit that we do not investigate here is that the fact that the product of reserves of the AMMs using GMM does not decrease and strictly increases with convergent orders should mean that their liquidity pools grow and thus reduce slippage and improve terms of trade for future traders.} To do this exercise we first note a peculiar feature of arbitrage. Exploiting all arbitrage opportunities when all AMMs use the CPMM mean that the liquidity ratios must be equal across AMMs. The common ratio, however, may vary depending on how arbitrageurs extract their profit. For instance, if arbitrageurs structure their swaps to realize profits exclusively in $Y$, ensuring that the net position of asset $X$ across all arbitrage swaps remains zero, the global liquidity ratio $\frac{\sum_i y_i}{\sum_i x_i}$ will decrease, leading to a corresponding decline in the common individual liquidity ratios $\frac{y_i}{x_i}$. To avoid contemplating all possible ways of resolving arbitrage opportunities, we focus on the most elementary case in which arbitrage profit extraction preserves the global liquidity ratio, ensuring that the resulting individual liquidity ratios remain equal to the original global liquidity ratio. We refer to this assumption as {\it balanced arbitrage}.

\begin{proposition}
Consider an initial distribution of liquidity $(\vec{x},\vec{y})$ and a swap $\Delta x > 0$ of $X$ that is optimally routed and compare two scenarios: 
\begin{enumerate}
\item All AMMs use the CPMM, and all strictly profitable arbitrage opportunities in the initial liquidity distribution are exploited by balanced arbitrage before the swap $\Delta x$ is executed.  
\item All AMMs use the GMM algorithm.  
\end{enumerate}
 (a) provides the trader strictly better terms of trade than (b) if and only if:
    \begin{equation}
        \frac{r}{1+\frac{\Delta x}{x}} > \frac{r_i}{1+\frac{\Delta x}{x_i}} < \frac{ r}{ 1+\frac{\Delta x}{\sqrt{\frac{ \max\{x_jy_j\}_{j\in I}}{r} }}} , \, \forall i \in I.
        \label{eq:tradersimpr}
    \end{equation} 
    \label{pro:tradersprofits}
\end{proposition}

\begin{proof}
We begin with some remarks that apply to both parts of the proof. In (a), the CPMM mechanism and balanced arbitrage imply that AMM $j$'s reserves $(x_j',y_j')$ satisfy the equations:
\begin{eqnarray*}
    x_j' y_j' &=& x_j y_j, \\
    \frac{x_j'}{y_j'} &=& r,
\end{eqnarray*}
which yield the solutions: $x_j'=\sqrt{\frac{x_jy_j}{r} }$ and $y_j'=\sqrt{r x_jy_j }$. Consequently, the terms of trade for a swap $\Delta x>0$ sent to AMM $j$ are:
\begin{equation}
    \frac{ \sqrt{r x_jy_j } }{ \sqrt{\frac{ x_jy_j}{r} } +\Delta x}=\frac{ r}{ 1+\frac{\Delta x}{\sqrt{\frac{ x_jy_j}{r} }}},
    \label{eqCPMMproof}
\end{equation}
where the optimal strategy for the trader is to swap with the AMM $j$ that maximizes $x_jy_j$. This implies that the last term of \eqref{eq:tradersimpr} represents the most favorable terms of trade for the trader in (a).

In (b), the terms of trade for a swap $\Delta x$ sent to AMM $i$ are:
\begin{equation}
\min \left\{ \frac{ r}{ 1+\frac{\Delta x}{x }}, \frac{ r_i}{ 1+\frac{\Delta x}{x_i }} \right\}.    
\label{eqGMMproof}
\end{equation}

We now use \eqref{eqCPMMproof} and \eqref{eqGMMproof} to prove the necessary and sufficient conditions, beginning with the "if" direction. The first inequality in \eqref{eq:tradersimpr} implies that the terms of trade in  (b) are determined by the second term in \eqref{eqGMMproof}. Combining this with the second inequality in \eqref{eq:tradersimpr} and \eqref{eqCPMMproof} yields the desired result.

For the "only if" direction, we proceed by contradiction starting with the first inequality in (\ref{eq:tradersimpr}). Suppose there exists some $i \in I$ such that:
\[
\frac{r}{1+\frac{\Delta x}{x}} \leq  \frac{r_i}{1+\frac{\Delta x}{x_i}}.
\]
This inequality, along with \eqref{eqCPMMproof} and \eqref{eqGMMproof}, implies that to complete the contradiction argument, it suffices to show:
\[
x \geq \sqrt{\frac{x_jy_j}{r}} \hspace{.5cm} \forall j\in I,
\]
To check that this inequality holds true it is replace $r = \frac{y}{x}$ to get:
\[
\sqrt{xy} \geq \sqrt{x_jy_j} \hspace{.5cm} \forall j\in I,
\]
which is clearly satisfied as desired.  

To conclude the proof, suppose now that 
\begin{equation*}
        \frac{r}{1+\frac{\Delta x}{x}} > \frac{r_i}{1+\frac{\Delta x}{x_i}} \geq \frac{ r}{ 1+\frac{\Delta x}{\sqrt{\frac{ \max\{x_jy_j\}_{j\in I}}{r} }}} , \, \forall i \in I.
    \end{equation*} 
The first inequality and \eqref{eqGMMproof}, mean that the most convenient terms of trade for the trader in (b) are:
\[\max_{i\in I} \frac{r_i}{1+\frac{\Delta x}{x_i}}, \]
which by the second inequality \eqref{eqCPMMproof} are more beneficial to the trader than the most convenient terms of trade in (a), as desired.
\end{proof}

To provide an intuitive interpretation of the Proposition, note that, in both the CPMM and GMM algorithms, the terms of trade depend on the ratio of reserves and slippage. For instance, in the case of the CPMM, when there is no arbitrage or non-convergent GMM orders, a swap $\Delta x$ delivers:
\[
\frac{r_i}{1+\frac{\Delta x}{r_i}}\Delta x.
\]
In the case of convergent orders under the GMM (priced as in the nGMM algorithm), the swap delivers:
\[
\frac{r}{1+\frac{\Delta x}{r}}\Delta x.
\]
Similarly, one can show that the CPMM after balanced arbitrage delivers:
\[
\frac{r}{1+\frac{\Delta x}{\sqrt{\frac{x_jy_j}{r}}}}\Delta x.
\]

Thus, the first inequality in \eqref{eq:tradersimpr} requires that all AMMs price GMM orders using CPMM prices, meaning there are no possible convergent swaps for $\Delta x$. If this holds, the second inequality ensures that GMM prices are worse for traders than CPMM prices with arbitrage. This occurs because arbitrage improves the best CPMM price for traders. Clearly, while these conditions can be met in reality, they are very demanding. 

The conditions will not hold in the following cases:

\begin{itemize}
    \item All AMMs have the same reserve ratio.  
    In this case, there is no room for arbitrage in (a) and no convergent orders in (b), meaning there is no difference between the prices offered by the GMM and the CPMM.\footnote{Formally, if $r_i = r$, then $\sqrt{\frac{x_jy_j}{r}}=\sqrt{\frac{x_jy_j}{r_j}}=x_j$, which violates the second inequality in \eqref{eq:tradersimpr} for any $i$.}
    
    \item All AMMs have different reserve ratios, but $\Delta x$ is small.  
    Heterogeneous reserve ratios and a small order size mean that some AMMs will display a convergent swap in (b). But convergent swaps always offer better prices to traders than CPMM prices with balanced arbitrage: the reserve ratios remain the same, but slippage is lower because the GMM uses as a reference aggregate reserves.\footnote{In the limit as $\Delta x \to 0$, \eqref{eq:tradersimpr} fails, as it converges to $r > r_i < r$ for all $i \in I$.}
    
    \item All AMMs have the same product of reserves, i.e., $x_jy_j$ is constant across $j$. Again, convergent swaps offer better prices to traders than CPMM prices with balanced arbitrage, so (a) can be better only when (b)´s best prices for traders are CPMM's prices. But in this case and in both scenarios, AMMs differences can only arise because of differences in reserve ratios. Arbitrage eliminates them in (a), whereas the heterogeneity of reserve ratios in (b) allows traders to find better deals.\footnote{A constant product $x_jy_j$ means that for any $l \in I$, $\sqrt{\frac{\max\{x_jy_j\}_{j\in I}}{r}}= \sqrt{\frac{x_ly_l}{r}}= x_l \sqrt{\frac{r_l}{r}}$, which is strictly less than $x_l$ for any $r_l<r$. Thus, the second inequality in \eqref{eq:tradersimpr} fails.}
\end{itemize}  

Still Proposition \ref{pro:tradersprofits} seems to say that the GMM is not trade preserving. We are now going to consider this question more in detail. Note that the comparison in the definition of trade preserving requires that all arbitrage opportunities are exploited, and this is not contemplated in scenario (b) of Proposition \ref{pro:tradersprofits}. Proposition \ref{cor:arbitrage} seems to suggest that this is irrelevant, but it is not difficult to show that the GMM allows for zero profit arbitrage opportunities, and thus a marginal decrease of the GMM prices in the convergent zone will make them strictly profitable. It may be shows that this is sufficient to make this marginally modified GMM trade preserving. However, we believe that there is more interesting solution. Keep the GMM algorithm as it is and include in the GMM an additional algorithm that implements the zero profit arbitrage opportunities required to make the GMM trade preserving. This is what we do in the next section.

\subsection{The GMM with Rebalancing}

In this section, we shall show how to modify the GMM algorithm to implement automatically all the zero profit arbitrage trades necessary (and no more than them) to make it trade preserving. 

Clearly, this require defining how the terms of trade of swaps between AMMs using the GMM algorithm are priced. All our results hold true if these swaps are implemented by an external intermediary, but in practical terms it may be more convenient to do so directly. Note that this is a feature that does not require special consideration for the CPMM prices. However the nGMM prices exhibit a particular property: any swap between AMMs do not affect the global aggregate $(x,y)$. Thus, for any swap $\Delta x$ of $X$ sent from one AMM to another one, any corresponding $\Delta y$ is compatible with maintaining a constant aggregate product. 

Here, we adopt the natural exchange rate given by the ratio of reserves, $\frac{y}{x}$, so that $\Delta y_{nGMM} = \frac{y}{x} \Delta x$ ensures that each AMM preserves its value, computed at the current aggregate marginal prices, i.e.,
\[
 (y_i-\Delta y_{nGMM}) + \frac{y}{x} (x_i+\Delta x) = (y_i-\frac{y}{x} \Delta x) + \frac{y}{x} (x_i+\Delta x)=y_i+\frac{y}{x} x_i.
\]
Once the nGMM is defined for these inter-AMM swaps, the GMM can be defined as in Section \ref{sec:GMM}, and the properties discussed there apply in the same manner to these swaps. Indeed, as we shall show later, these swaps are equivalent to other swaps with zero net trade implemented by an external agent. From now on, we assume that the GMM incorporates these specific pricing rules for inter-AMM trades.

Next, we introduce the modification of the GMM that guarantees that trade preservation.

\begin{definition}
    The GMM algorithm with rebalancing is an algorithm that applies the GMM algorithm to the resulting updated vector of reserves from executing the following algorithm iteratively until one of the conditions fails (assume wlog $\Delta x>0)$.
    \begin{enumerate}
        \item Check whether $l\in \arg \max\{x_jy_j\}_{j\in I}$.
        \item Check whether \eqref{eq:tradersimpr} is met.
        \item Check whether $\frac{\tilde y_l}{\tilde x_l} < r$, for the (updated) vector of reserves $(\tilde x_l,\tilde y_l)$.
        \item Update reserves using a swap from AMM $l$ to $j\in \arg \max_{k \in I\setminus \{l\}} \frac{\tilde{y}_k}{\tilde x_k}$ of \[\min \left\{ \frac{r\tilde{x}_j-\tilde{y}_j}{2r},\frac{\tilde{y}_l-r\tilde{x}_l}{2r} \right\}\] units of $X$.

        \item Go to (c).
    \end{enumerate}

    \end{definition}

The GMM algorithm with rebalancing is a variation of the GMM algorithm that, before applying it, checks whether \eqref{eq:tradersimpr} holds. It is under this condition that Proposition \ref{pro:tradersprofits} states that the AMM provides better terms of trade for the trader under (a) than (b). Specifically, the trader benefits from trading with the AMM that has the largest product of reserves (and thus the lowest slippage) after all arbitrage opportunities are exhausted and all AMMs have a reserve ratio equal to the global ratio.

In this case, the GMM algorithm with rebalancing redistributes swaps to the other AMMs until the AMM's reserve ratio aligns with the global ratio. Consequently, the GMM with rebalancing offers marginal terms of trade at least as favorable as those in the CPMM with arbitrage. Additionally, slippage is lower since rebalancing increases reserves because the AMM captures a portion of the arbitrageurs' profits.

\begin{proposition}
    Consider an initial distribution of liquidity $(\vec{x},\vec{y})$ and a swap $\Delta x > 0$ of $X$ (wlog) that is optimally routed. Compare the following two scenarios:
    
    \begin{enumerate}
        \item All AMMs use the CPMM, and all strictly profitable arbitrage opportunities in the initial liquidity distribution are exploited by balanced arbitrage before the swap $\Delta x$ is executed.
        \item All AMMs use the GMM algorithm with rebalancing.
    \end{enumerate}
    
    (b) provides better terms of trade for the trader than (a).
\end{proposition}

\begin{proof}
Proposition \ref{pro:tradersprofits} and the definition of GMM with rebalancing means that we can restrict to the case where \eqref{eq:tradersimpr} is satisfied. Suppose from now on that this is the case. In (a), arbitrage equalises the ratio of reserves of all AMMs. Solving that the product of reserves of each AMM remains constant and that each ratio of reserves must equal to $r$, one can use the CPMM formula to show that AMM $j$ would respond to the swap of $\Delta x$ sending back 
\begin{equation}
    \frac{ r}{ 1+\frac{\Delta x}{\sqrt{\frac{ x_jy_j}{r} }}},
    \label{eq:CPMMsc1}
\end{equation}
so that the most profitable trade for the trader is the AMM with largest product of reserves, say $l$. To prove the proposition we shall show that the same AMM in (b), offers strictly better terms of trade. Since $l=\arg \max_{j\in I} x_jy_j$, the second inequality of \eqref{eq:tradersimpr} implies that 
\begin{equation}
     \frac{r_l}{1+\frac{\Delta x}{x_l}} < \frac{ r}{ 1+\frac{\Delta x}{\sqrt{\frac{ x_ly_l}{r} }}}=\frac{ r}{ 1+\sqrt{\frac{ r}{r_l} \frac{\Delta x}{x_l }}},
     \label{eq:proofGMMcomp}
\end{equation}
    which implies that $\frac{y_l}{x_l}=r_l<r$, and thus there are other AMMs with ratios of reserves larger than $r$. The algorithm GMM with rebalancing iterates sending swaps from $l$ to each of other AMMs in step 4. At each iteration in which the updated ratio of reserves of $l$ and $j$, $\tilde r_l\equiv \frac{\tilde y_l}{\tilde x_l}$ and $\tilde r_l\equiv \frac{\tilde y_j}{\tilde x_j}$ respectively, satisfy $\tilde r_l<r<\tilde r_j$, there are two possibilities. Consider first that,
    \begin{equation}
        \frac{r\tilde x_l-\tilde y_l}{2r} > \frac{\tilde y_j-r\tilde x_j}{2r},
        \label{eq:ineq}
    \end{equation}
    then the swap sent to $j$ is equal to $\tilde \Delta x=\frac{\tilde y_j- r\tilde x_j}{2r}$ units of $X$. Applying the GMM with rebalancing formula, the amount of $Y$ obtained with the swap  by $l$ is equal to:
    \begin{eqnarray*}\min \left\{ \frac{\tilde y_j}{\tilde x_j+\tilde \Delta x},r\right\}\tilde \Delta x&=&\min \left\{ \frac{\tilde y_j}{\tilde x_j+\frac{\tilde y_j-r\tilde x_j}{2r}},r\right\}\tilde \Delta x\\
    &=&\min \left\{ \frac{\tilde r_j}{1+\frac{\tilde r_j-r}{2r}},r\right\}\tilde \Delta x\\
    &=&\min \left\{ \frac{\tilde r_j}{r+r_j},1\right\}r\tilde \Delta x\\
    &=& r \tilde \Delta x.
    \end{eqnarray*}
    Thus, the swap sent to $j$ increases the ratio of reserves of $l$ to:
    \[\frac{\tilde y_l+r \tilde \Delta x}{\tilde x_l-\tilde \Delta x}=\frac{\tilde y_l+\frac{\tilde y_j-r\tilde x_j}{2}}{\tilde x_l-\frac{\tilde y_j-r\tilde x_j}{2r}} <\frac{\tilde y_l+\frac{ r \tilde x_l-\tilde y_l }{2}}{\tilde x_l-\frac{r \tilde x_l-\tilde y_l }{2r}}=r .\]
    Thus, the algorithm goes back to 3. and then 4. again, repeating the process. Note that there are other AMMs with reserve ratios strictly larger than $r$ since the updated ratio of reserves of $l$ is strictly less than $r$ and $r$ is the aggregate ratio of reserves.
Consider now the case:
    \begin{equation}
        \frac{r\tilde x_l-\tilde y_l}{2r} \leq \frac{\tilde y_j-r\tilde x_j}{2r}.
        \label{eq:ineq2}
    \end{equation}
    A similar argument as above implies that the ratio of reserves of $l$ increases to $r$ and then the algorithm stops at 3. once it revisits it again. Since this is the only way the iterative part of the algorithm stops. The amount of $Y$ that the algorithm returns to the trader must be the minimum between the nGMM quantity and the CPMM quantity that corresponds to the updated liquidity reserves. The former is equal to:
    \begin{equation}
\frac{y}{x+\Delta x}\Delta x=\frac{r}{1+\frac{\Delta x}{x}}\Delta x,
        \label{eqGMMproof2}
    \end{equation}
    and the latter to:
    \begin{equation}
    \frac{\tilde y_l}{\tilde x_l+\Delta x}\Delta x=\frac{r}{1+\frac{\Delta}{\tilde x_l}}\Delta x=\frac{r}{1+\frac{\Delta}{x_l\left(\frac{r+r_l}{2r}\right)}}\Delta x,
        \label{eqCPMMproof2}
    \end{equation}
       where $(\tilde x_l,\tilde y_l)$ denotes the final vector of liquidity reserves of $l$ after running the iterative part of the algorithm. Since this part only stops at the point in which $\frac{\tilde y_l}{\tilde x_l}=r$ and it satisfies $r(x_l-\tilde x_l)=y_l-\tilde x_l$ we have that $(\tilde x_l,\tilde y_l)=\left( x_l\frac{r+r_l}{2r},x_l\frac{r+r_l}{2}\right)$. Since $x_l\frac{r+r_l}{2r}\leq x_l\leq x$, \eqref{eqCPMMproof2} is less than \eqref{eqGMMproof2}, and thus the amount of $\Delta y$ returned by the algorithm is equal to \eqref{eqCPMMproof2}. To finish the proof we show that \eqref{eqCPMMproof2} is greater than \eqref{eq:CPMMsc1} for $j=l$. This is equivalent to show that $\frac{r+r_l}{2r}\geq \sqrt{\frac{r_l}{r}}$, which can be easily checked.
\end{proof}

One concern, however, is that rebalancing may introduce perverse effects. We argue that this is not the case by showing that rebalancing produces the same outcome for the AMMs as a specific sequence of trades with zero net trade applied to the standard GMM. For simplicity, we illustrate this claim with an example, though the argument is general.

\begin{examplebox}{Toy example - Part 7: Rebalancing} 

    Suppose a set of two AMMs with initial vector of reserves \[(\vec{x},\vec{y})=(90ETH,210ETH,440000UST,760000UST), \] and a swap of 1 ETH sent by a trader to AMM (b). In this case, the GMM with balancing applies the iterative algorithm sending 10ETH to AMM 1 to get 40000UST in return ($\frac{1200000 UST}{300ETH}*10ETH$). Then, the vector of reserves is updated to \[(100ETH,200ETH,400000UST,800000UST),\] and AMM 2 returns to the trader $3980,10 UST \approx \frac{800000UST}{200+1}ETH*1ETH$ since it is a divergent order.

    Consider next, the standard GMM and that the swap sent by the trader is preceded by the following two trades sent by an additional trader. The additional trader sends 10ETH to AMM1 to get 38709.67 UST in return ($\approx \frac{1200000 UST}{300+10 ETH}*10ETH$ since it is a convergent trade). This updates the vector of reserves to \[(100ETH,210ETH,436.129,03UST,760000UST),\] and then sends 38709.67 UST to AMM2 to get 10ETH in return ($\approx \frac{310ETH * 38709.67 UST}{1.161.290,32 + 38709.67 UST }  $ since it is again a convergent trade). After these two trades, the vector of reserves is updated to \[(100ETH,200ETH,400000UST,800000UST) ,\] and now when the original trade is submitted to AMM 2 it returns to the trader $3980,10 UST \approx \frac{800000UST}{200+1}ETH*1ETH$ since it is a divergent order.

\end{examplebox}

Thus, rebalancing neither enables additional manipulative strategies nor alters the cost of manipulation compared to using phony trades. The only caveat is that manipulation with phony trades requires additional asset holdings (e.g., 10 ETH in the example).

\begin{corollary}
The GMM with rebalancing is trade preserving and not exploitable.
\end{corollary}


\subsection{MEV Sandwich Attacks}

Next, we show MEV sandwich are strictly less profitable when the AMM uses the GMM algorithm than when it uses the CPMM.

\begin{definition}
    Given a swap of $\Delta x>0$ units of $X$ submitted to AMM $i$, a MEV sandwich attack of $\widehat{\Delta x}$ units of $X$ consists of a frontrunning swap and a backrunning swap submitted to AMM $i$ before and after $\Delta x$, respectively. The frontrunning swap sends $\widehat{\Delta x}$ units of $X$ and the backrunning swap sends the units of $Y$ obtained with the first swap.
\end{definition}

A MEV sandwich may be profitable due to slippage: the first order buys $X$ more cheaply than the price at which the second order sells because there is an order in between that makes $X$ more scarce at the AMM. This is clearly the case when the AMM uses CPMM in which prices reflect the scarcity at the AMM but it is less so in the case of the GMM. In this later case, the first buy order is satisfied at a price that reflects the scarcity created by the order but the subsequent sell order is satisfied at a price that reflects the much smaller scarcity existing in the set of AMMs. The reason for this is that one would expect that AMMs should start from an equilibrium situation in which all (marginal) prices are the same (and thus local levels of scarcity) and thus the first buy order moves the AMM away from the other AMMs whereas the sell order moves the AMM towards the other AMMs. Next proposition formalises this argument. 

\begin{proposition}
    The profit of an elementary MEV sandwich is strictly lower when the AMM uses GMM than when the AMM uses CPMM.
    \label{proMEVGMM1}
\end{proposition}

\begin{proof} By definition, the profit of an elementary MEV sandwich is equal to $\widehat{\Delta^F x}_s-\widehat{\Delta x},$

for algorithm $s\in \{CPMM,GMM\}$, where:
\begin{eqnarray*}
    \widehat{\Delta} y_{CPMM}&=&\frac{y_i}{x_i+\widehat \Delta x}\widehat \Delta x,\\
    \Delta y_{CPMM}&=&\frac{y_i-\widehat{\Delta} y_{CPMM}}{x_i+\widehat \Delta x+\Delta x} \Delta x,\\
    \widehat \Delta^F x_{CPMM}&=& \frac{x_i+\widehat \Delta x+\Delta x}{y_i- \Delta y_{CPMM}} \widehat \Delta y_{CPMM},
\end{eqnarray*}
and:
\begin{eqnarray*}
    \widehat{\Delta} y_{GMM}&=&\min \left\{ \frac{y_i}{x_i+\widehat \Delta x},\frac{y}{x+\widehat \Delta x}\right\} \widehat \Delta x\\
    \Delta y_{GMM}&=&\min\left\{ \frac{y_i-\widehat{\Delta} y_{GMM}}{x_i+\widehat \Delta x+\Delta x},\frac{y-\widehat{\Delta} y_{GMM}}{x+\widehat \Delta x+\Delta x}\right\} \Delta x,\\
    \widehat \Delta^F x_{GMM}&=& \min\left\{ \frac{x_i+\widehat \Delta x+\Delta x}{y_i-\Delta y_{GMM}} , \frac{x+\widehat \Delta x+\Delta x}{y-\Delta y_{GMM}} \right\} \widehat \Delta y_{GMM}.
\end{eqnarray*}
Thus, the min implies that the profits are weakly less under the GMM than under the CPMM. To show that they are strictly less we only need to show that in one of the mins, the second term is strictly less then the first. For that, just note that if it were not the first one, the following chain of implications means that it has to be the third one:
\begin{eqnarray*}
    \frac{y_i}{x_i+\widehat \Delta x} &\leq &  \frac{y}{x+\widehat \Delta x}\\
    \Rightarrow \frac{y_i}{x_i+\widehat \Delta x+\Delta x} &< & \frac{y}{x+\widehat \Delta x+\Delta x}\\
    \Rightarrow \frac{x_i+\widehat \Delta x+\Delta x}{y_i} &>& \frac{x+\widehat \Delta x+\Delta x}{y}\\
    \Rightarrow \frac{x_i+\widehat \Delta x+\Delta x}{y_i-\Delta y_{GMM}} &>& \frac{x+\widehat \Delta x+\Delta x}{y- \Delta y_{GMM}}.
\end{eqnarray*}
The first line says that in the first min, the left term is weakly less than the right one; the last line says that in the third min, the left term is strictly greater than the right one; the second implication is direct and the first one and the last one follow from the observation that $y_i(x+\beta)-y(x_i+\beta)$ it is strictly decreasing in $\beta$ for $y_i<y$.  
\end{proof}

Next, we redo the MEV extraction example computed for CPMM in Example 2 in the case of GMM and we observe that the extracted MEV is so small that it can become negative.

\begin{examplebox}{Toy example - Part 7: MEV extraction in GMM} 

    Consider two AMMs using the GMM algoritm and with a liquidity pool containing 100 ETH and 400,000 UST each. The marginal price of ETH is then 4,000 UST/ETH. Suppose that a trader interested in converting 40,000 UST in ETH sends them to AMM 1. The MEV-extractor can perform a sandwich attack by sending 60,000 UST to AMM 1 just before the victim's transaction is executed and to send the proceeds back to AMM 1 immediately after. This would work as follows. Note that along all trades the reserves of AMM 2 remain constant and equal to 100 ETH and 400,000 UST.

\begin{itemize}
    
\item \textbf{Frontrunning:} MEV-extractor transaction 1 sends 60,000 UST to AMM 1 to get 13.0435 ETH in return since this order is divergent and thus priced as the CPMM. This order changes the vector of reserves of AMM 1 to 86,9565 ETH and 460.000 UST. Its marginal price of ETH increases to 5,290 UST/ETH.

\item \textbf{Victim's transaction:} The victim sends 40,000 UST to AMM 1. This order is also divergent and thus priced as in the CPMM. The trader obtains in return 6.9565 ETH. The resulting reserves of AMM 1 are then 80 ETH and 500,000 UST. Nothing changes in AMM 2.

\item \textbf{Backrunning:} MEV-extractor transaction 2 sends 13.0435 ETH to AMM 1. In this case, the order is convergent and thus priced by the nGMM. Since the global reserves are 180 ETH and 900,000 UST, the MEV-extractor gets 60,811 UST. The MEV-extractor makes a net profit of just 811 UST, substantially less than the profit of 10,094 UST obtained in the CPMM.

\end{itemize} 
      
\end{examplebox}

We also provide a general analysis of the profits of an elementary MEV sandwich. In this case, we also need to specify the global reserves. To simplify the notation we assume that the ratio of global reserves $\frac{t}{x}$ is equal to the ratio of reserves of AMM i $\frac{y_i}{x_i}$. 

\begin{proposition}
    The profit of an elementary MEV sandwich $\hat \Delta x$ to a trader order $\Delta x$ to an AMM i with liquidity reserves $(x_i,r x_i)$ and global reserves $(x,r x)$ running the GMM algorithm is equal to:
    \begin{equation}
        \left( \frac{ \left(1+ \frac{\hat \Delta x}{x}+\frac{ \Delta x}{x} \right) \left(1+ \frac{\hat \Delta x}{x_i}+\frac{ \Delta x}{x_i} \right)}{  \left(1+ \frac{\hat \Delta x}{x_i}+\frac{ \Delta x}{x_i} \right)\left(1+ \frac{\hat \Delta x}{x_i} \right)-\frac{\Delta x}{x}} -1 \right) \hat \Delta x
    \end{equation}
\end{proposition}

\begin{proof}
As in the proof of Proposition \ref{proMEVGMM1}, we can compute the profits of MEV extractor as $\hat \Delta^F x-\hat \Delta x$ where:
\begin{equation*}
\begin{array}{rcl}
    \widehat{\Delta} y = & \min \left\{ \frac{r}{1+\frac{\widehat \Delta x}{x_i}},\frac{r}{1+\frac{\widehat \Delta x}{x}}\right\} \widehat \Delta x 
    &= \frac{r}{1+\frac{\widehat \Delta x}{x_i}} \widehat \Delta x,\\
    \Delta y =& \min\left\{ \frac{r-\frac{\widehat{\Delta} y}{x_i}}{1+\frac{\widehat \Delta x}{x_i} +\frac{ \Delta x}{x_i}},\frac{r-\frac{\widehat{\Delta} y}{x}}{1+\frac{\widehat \Delta x}{x} +\frac{ \Delta x}{x}}\right\} \Delta x 
    &= \frac{r-\frac{\widehat{\Delta} y}{x_i}}{1+\frac{\widehat \Delta x}{x_i} +\frac{ \Delta x}{x_i}} \Delta x,\\
    \widehat \Delta^F x =& \min\left\{ \frac{1+\frac{\widehat \Delta x}{x_i} +\frac{\Delta x}{x_i}}{r-\frac{\Delta y}{x_i}} , \frac{1+\frac{\widehat \Delta x}{x} +\frac{\Delta x}{x}}{r-\frac{\Delta y}{x}} \right\} \widehat \Delta y 
    &= \frac{1+\frac{\widehat \Delta x}{x} +\frac{\Delta x}{x}}{r-\frac{\Delta y}{x}} \widehat \Delta y.
\end{array}
\end{equation*}
We can derive the proposition from the recursive substitution of $\Delta y $ and $\widehat \Delta y$ and some straightforward simplifications.
\end{proof}

\begin{figure}[ht!]
    \centering
    \includegraphics[width=10cm]{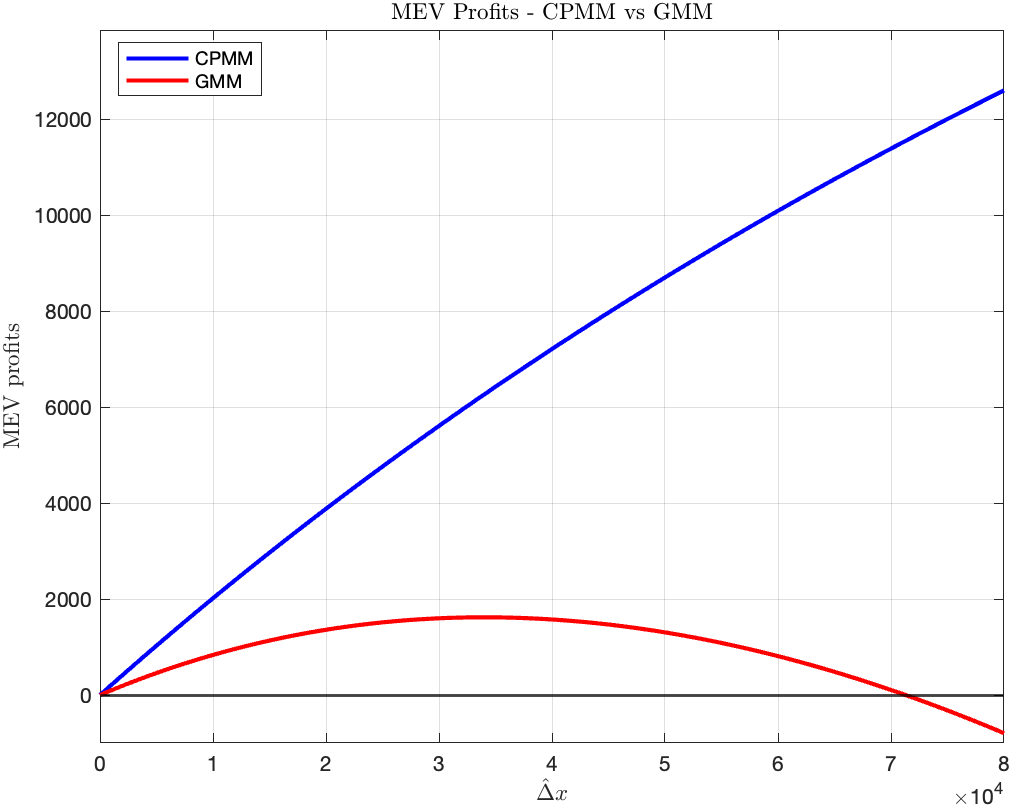} 
    \caption{The graph plots the profits of elementary MEV sandwich manipulations $\hat \Delta x$ for $x_i=400,000$, $x=800.000$ and $\Delta x =40,000$.}
    \label{fig:MEV_GMM}
\end{figure}

Figure \ref{fig:MEV_GMM} illustrates the proposition for different values of $\hat \Delta x$ and compare the results with MEV profits obtained in the CPMM case. The GMM reduces substantially the profitability of MEV attacks and even render them unprofitable.

\subsection{Impermanent Losses}

In the case of impermanent losses for the GMM we need to keep track how traders route their orders as the price changes from $r_0$ to $r$ since unlike in the CPMM, this affects how the value of the reserves of the AMM evolve.

We adopt a simple setting that we refer as the {\it ideal benchmark}. In the ideal benchmark, we assume that there are two AMMs only $1$ and $2$, with respectively reserves $(x_1,y_1)$ and $(x_2,y_2)$. 
As in the case of the CPMM, we also assume that the initial ratio of reserves of both AMMs is equal to the initial price $r'$, i.e. $\frac{y_1}{x_1}=\frac{y_2}{x_2}=r'$ and we model a change in market sentiment assuming an insider trader that submit trades to the AMM to maximize the value of its net final position at a the new price $r$. 

Note that the impermanent losses under the ideal benchmark are an upper bound as adding additional trades can only increase the value of the final liquidity reserves of the AMMs.

Finally, we assume without loss of generality that $\alpha\equiv \frac{x_2}{x_1+x_2} \leq 0.5$, which under the above assumptions means that AMM 1 is the AMM with largest initial market value $ r' x_i +y_i$.

\begin{proposition}
    In the ideal benchmark, the impermanent losses of AMM 1 are the same as in the case of the CPMM, whereas the impermament losses of AMM 2 are equal to:
    \begin{equation}
        IL_2=1-2\frac{\sqrt{\left(\sqrt{\frac{r'}{r}}+\frac{1-\alpha}{\alpha} \right) \left(\sqrt{\frac{r}{r'}}+ \frac{1-\alpha}{\alpha} \right) } -\frac{1-\alpha}{\alpha} }{\sqrt{\frac{r'}{r}}+\sqrt{\frac{r}{r'}}}.
        \label{eq:ILGMM}
    \end{equation}
    \label{proILGMM}
\end{proposition}

\begin{proof}
Assume $r'<r$. The other case is symmetric. Since AMM 1 slippage is lower but marginal prices are the same, it is optimal for the insider trader to trade with AMM 1 first. This first order is divergent and thus priced by CPMM prices, so the quantity $\Delta x^*_1$ of $X$ sent to AMM 1 solves:
\[\max_{\Delta x_1} \left( \frac{r}{1+\frac{\Delta x_1}{x_1}}  - r' \right) \Delta x_1,\]
this is $\Delta x^*_1=  x_1 \left( \sqrt{\frac{r}{r'}}-1 \right)$, and thus,
\[\Delta y^*_1\equiv \frac{r}{1+\frac{\Delta x^*_1}{x_1}}\Delta x_1^*,\] is the quantity of $Y$ received by the insider trader. Note that the optimal value $\Delta x_1^*$ is such that the updated ratio of reserves of AMM 1 is equal to the new price $r$: 
\begin{equation}
    \frac{y_1-\Delta y^*_1}{x_1+\Delta x^*_1}=\frac{r-\frac{r}{1+\frac{\Delta x^*_1}{x_1}}\frac{\Delta x^*_1}{x_1}}{1+\frac{\Delta x^*_1}{x_1}}=\frac{r}{\left(1+\frac{\Delta x^*_1}{x_1}\right)^2}=r',
    \label{eqrprime1}
\end{equation}
which means that $(x_1+\Delta x^*_1)(y_1-\Delta y^*_1)=r' (x_1+\Delta x^*_1)^2$, and thus that,
\begin{equation}
    x_1=\sqrt{\frac{r'}{r}}\left(x_1+\Delta x^*_1 \right),
    \label{eqx1r}
\end{equation} 
since $(x_1+\Delta x^*_1)(y_1-\Delta y^*_1)=x_1y_1=rx_1^2$.

The second optimal order for the insider trader is to send to AMM 2, the convergent order $\Delta x^*_2$ of $X$ that solves:
\[\max_{\Delta x_2} \left( \frac{\hat r}{1+\frac{\Delta x_2}{x^U}}  - r' \right) \Delta x_2,\]
and returns an amount of $Y$ that we denote by $\Delta y^*_2$. By a similar argument as for $\Delta x_1^*$:
\begin{equation}
    \frac{x_2+\Delta x_2^*+x_1+\Delta x_1^*}{y_2-\Delta y_2^*+y_1-\Delta y_1^*} =r',
    \label{eqrprime2}
\end{equation}
which together with \eqref{eqrprime1} means that:
\begin{equation}
    \frac{x_2+\Delta x_2^*}{y_2-\Delta y_2^*} =r'.
    \label{eqrprime3}
\end{equation}

Finally, the nGMM also means that:
\[(x_2+\Delta x_2^*+x_1+\Delta x_1^*)( y_2-\Delta y_2^*+y_1-\Delta y_1^*)=(x_2+x_1+\Delta x_1^*)( y_2+y_1-\Delta y_1^*),\]
which can be transformed into:
\[
    r'\left(x_2+\Delta x_2^*+\frac{1-\alpha}{\alpha} \sqrt{\frac{r}{r'}}x_2\right)^2=rx_2^2 \left(1+\frac{\alpha}{1-\alpha} \sqrt{\frac{r}{r'}}\right)\left(1+\frac{\alpha}{1-\alpha} \sqrt{\frac{r'}{r}}\right),
\]
using $x_1=\frac{1-\alpha}{\alpha} x_2$, $y_2=rx_2$, and \eqref{eqrprime1}-\eqref{eqrprime3}. Thus, solving for $x_2+\Delta x_2^*$ gives us:
\[x_2+\Delta x_2^*= x_2 \sqrt{\frac{r}{r'}} \left(\sqrt{ \left( 1+\frac{\alpha}{1-\alpha} \sqrt{\frac{r}{r'}}\right)\left(1+\frac{\alpha}{1-\alpha} \sqrt{\frac{r'}{r}}\right)}-\frac{1-\alpha}{\alpha}  \right).
\]
We use this expression to compute the value of AMM 2 at the final prices and reserves. 
\begin{eqnarray*}
    V_2 &\equiv& r' (x_2+\Delta x^*_2)+(y_2-\Delta y^*_2)\\
    &=& 2r'(x_2+\Delta x^*_2) \\
    &=& 2r'x_2 \sqrt{\frac{r}{r'}} \left(\sqrt{ \left( 1+\frac{\alpha}{1-\alpha} \sqrt{\frac{r}{r'}}\right)\left(1+\frac{\alpha}{1-\alpha} \sqrt{\frac{r'}{r}}\right)}-\frac{1-\alpha}{\alpha}  \right),
\end{eqnarray*}
whereas the value of the initial reserves of AMM 2 at the new price is:
\[V_2^0=r' x_2 +y_2=x_2(r'+r) .\]
which together with the definition of impermanent losses as $IL_2=1-\frac{V_2}{V^0_2}$ implies the proposition.
\end{proof}

\begin{figure}[ht!]
    \centering
    \includegraphics[width=10cm]{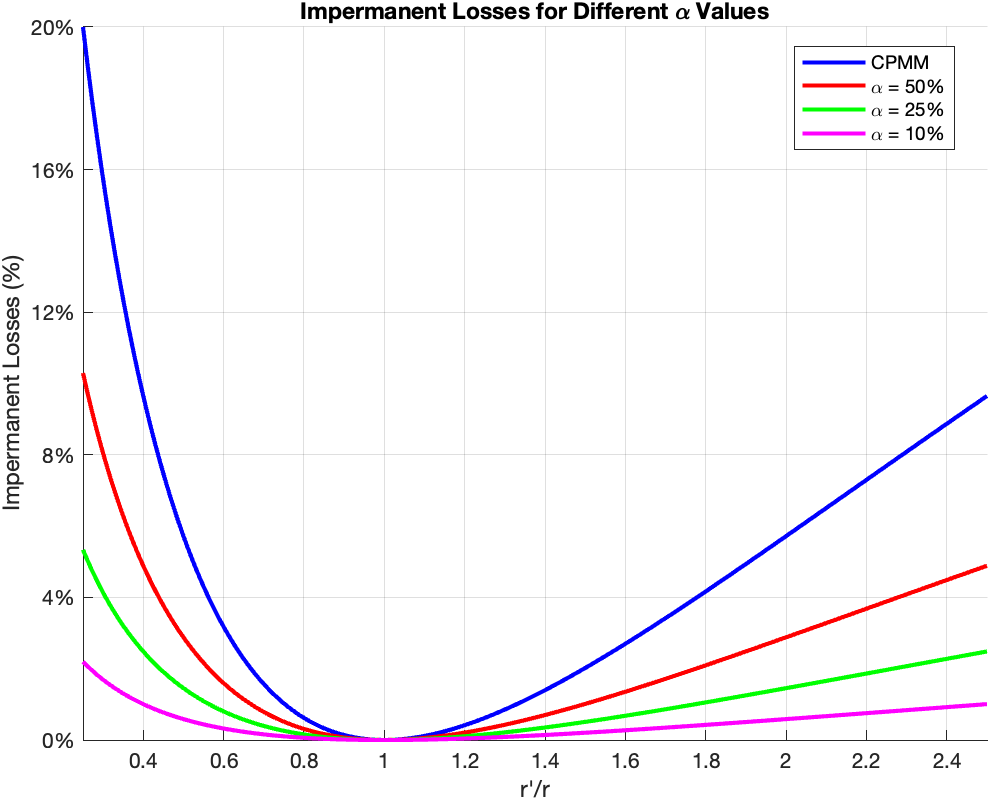} 
    \caption{The graph plots the impermanent losses as a function of the price variation $\frac{r}{r_o}$ for different values of $\alpha$ in the ideal benchmark using Equation \eqref{eq:ILGMM}.}
    \label{fig:GMM_IL}
\end{figure}

Figure \ref{fig:GMM_IL} illustrates Proposition \ref{proILGMM} and compares the impermananent losses in the GMM with respect to the CPMM. 

The next example illustrate this result.

\begin{examplebox}{Toy example - Part 8: Impermanent losses in GMM}     
        Consider two GMMs with liquidity pools containing 100 ETH and 400,00 UST. The marginal price is then 4,000 UST/ETH.
        
        Similarly than in Example 3, suppose that one or more traders submit transactions for buying 30 ETH
        on each GMM.
    
        We can compute impermanent loss IL) as the difference between the value of the GMM's original reserves valued at the current price and the current reserves valued at the current price. 
    
        To make a simple example, we assume that a first order for 30 ETH arrives to GMM1 and a second order for 30 ETH arrives to GMM2. The final reserves for GMM1 are the same as in one of the CPMMs in Example 3, namely 70 ETH and 571,429 UST.
    
        The final reserves of GMM2 resulting from the second trade of 30 ETH can be computed using the convergent formula for 28 ETH, at point which the relative prices are equalized for the two GMMs and the divergent formula for the remaining 2 ETH.
        
        Overall, this implies that the impermanent loss for GMM1 and GMM2 are equal to 129,023 UST, which is 17,899 UST less than in Example 3.         
\end{examplebox}

The next section provides simulations using real life transactions and provides a quantitative assessment of how large this reduction of impermanent losses is.

%% file: sections/Data.tex
We proceed to evaluate the relevance of our theoretical results using real data and simulations based on actual trades. We focus on transactions from 2023 in the most popular AMM on the Ethereum network, Uniswap V2, and uses the most elementary version of the CPMM algorithm.\footnote{In 2023, Uniswap V2 facilitated trades in 165,566 different token pairs, with a total trading volume of \$88 billion. In comparison, Ethereum’s total trading volume was \$513 billion, while the combined trading volume across all blockchains was \$885 billion. Thus, Uniswap V2 accounted for 17\% of Ethereum’s volume, and Ethereum processed 58\% of the total volume across all blockchains. Source: Dune Analytics \url{https://dune.com}.}${}^,$\footnote{Uniswap has produced more sophisticated version of the CPMM algorith called Uniswap v3 and Uniswap v4. They still use local reserves to determine the terms of trade.}
To construct our dataset, we extract Uniswap transactions from TheGraph, \url{https://thegraph.com}, and flag arbitrage and sandwich attacks on Ethereum using the identification provided by Zeromev, \url{https://zeromev.org/}, both ordered by block. We also obtain block-level WETH prices from Alchemy, \url{https://www.alchemy.com}. To express transaction values in U.S. dollars, we use the dollar prices of WETH and major stablecoins (DAI, USDC, and Tether), which allows us to convert approximately 90\% of the transaction volume. Pairs without reliable price data are excluded from the analysis.

Table~\ref{tab:summary} summarises the importance of arbitrage and sandwich attacks in our database relative to the fees charged by AMMs to traders. Since these fees are essentially the compensation to liquidity providers for locking their tokens in the AMM, they can be interpreted as the cost of liquidity provision. The table thus provides a comparison of the costs of arbitrage and sandwich attacks relative to the cost of liquidity provision. The table shows that arbitrage is not particularly significant, but sandwich attacks are.

\begin{table}[H]
\centering
\begin{tabular}{lc}
\toprule
\textbf{Metric} & \textbf{Value}  \\
\midrule
Fees & \$261m \\

Arbitrage profits & \$10m \\

Sandwich profits & \$440m \\

Arbitrage profits over fees & 3.8\% \\

Sandwich profits over fees & 168.2\% \\

\toprule
\textbf{Crypto pairs s.t.} & \textbf{\% of trade volume}  \\
\midrule
\vspace{.1cm}
$\tfrac{\mbox{Arbitrage profits}}{\mbox{Total fees}}> 50\%$ & 0.8\% \\
\vspace{.1cm}
$\tfrac{\mbox{Sandwich profits}}{\mbox{Total fees}}> 50\%$ & 41.8\% \\
\bottomrule
\end{tabular}
\caption{Fees, arbitrage profits, and sandwich profits for all pairs traded on Ethereum in 2023.}
\label{tab:summary}
\end{table}

Next, we use our dataset to simulate the profitability of the sandwich attacks recorded in 2023 under the hypothetical scenario in which Uniswap V2, instead of using the CPMM algorithm, employs the GMM algorithm. To this end, we focus on recorded sandwich attacks that are relatively simple to track, specifically those in which the frontrun trade is equal to the backrun trade. This includes elementary sandwich attacks but also more general cases, as we allow for several intermediate trades between the frontrun and backrun trades. This restriction reduces the observed sandwich profits from \$440 million to \$339 million.

One challenge in this exercise is tracking the reserves of the other AMMs that may be trading the same token pairs. Rather than relying on real-world reserve data, we run our algorithm under different hypothetical scenarios. In particular, we consider two broad cases. In the first, we assume that the reserves of the other AMMs are equal to~$\beta$ times the reserves of Uniswap v2, and we consider different values of~$\beta$. In this case, the attacker’s profits can be computed as:

    \begin{equation}
    \left( \frac{(1+\frac{\delta+\hat \delta}{1+\beta})(1+\delta+\hat \delta) }{\left( 1+\hat \delta \right)(1+\delta+\hat \delta)  -\frac{\delta}{1+\beta}  } -1 \right) \hat \Delta x,
    \end{equation}
    where $x_i$ denotes the reserves in Uniswap of the asset submitted by the attacker, $\hat{\Delta} x$ is the amount of this asset submitted by the attacker, $\Delta x$ is the total amount submitted by all traders, and $\delta \equiv \frac{\Delta x}{x_i}$ and $\hat{\delta} \equiv \frac{\hat{\Delta} x}{x_i}$.

The results are presented in Table~\ref{tab:beta}. We also include a column showing the percentage of cases in which the attacker’s profits decrease so significantly that they become negative. In the same table, we report the sandwich profits recorded by Zeromev and the simulated attacker profits under the CPMM algorithm (which corresponds to~$\beta = 0$). The table shows that our CPMM simulations produce results consistent with the data from Zeromev. It also shows that the attacker’s profits can be substantially reduced by switching from the CPMM to the GMM algorithm.

\begin{table}[H]
\centering
\begin{tabular}{lcc}
\toprule
\textbf{Algorithm} & \textbf{Attackers' profits} & \textbf{\% of profits$<$0} \\
\midrule
Zeromev     & \$338.70m & 4.1 \% \\
CPMM    & \$333.70m & 3.3 \% \\
GMM ($\beta=0.01$)  & \$255.72m & 6.7 \% \\
%
%
%
GMM ($\beta=0.05$)  & \$188.65m & 6.7 \% \\
GMM ($\beta=0.1$)   & \$164.40m & 6.8 \% \\
GMM ($\beta=0.5$)   & \$121.96m & 7.4 \% \\
%
GMM ($\beta=1$)     & \$107.22m & 7.9 \% \\
GMM ($\beta=1.5$)   & \$99.63m  & 8.2 \% \\
GMM ($\beta=10$)    & \$78.78m  & 9.5 \% \\
\bottomrule
\end{tabular}
\caption{Sandwiches reported by Zeromev and simulations for 2023: attacker's profits and percentage of cases with negative profits for different values of the fraction~$\beta$ of Uniswap V2's reserves held by other AMMs.}
\label{tab:beta}
\end{table}

In the second case, we conduct a different exercise. We consider a counterfactual scenario in which the Uniswap V2 pool of reserves is evenly divided among~$n$ AMMs, all using the GMM algorithm, and where each AMM computes global reserves by aggregating the reserves of all $n$ AMMs. We compute the attacker’s profits assuming that the attack and the trades had been directed to one of these hypothetical AMMs. A straightforward computation shows that these profits are given by:
\begin{equation}
\left( \frac{(1 + (\delta + \hat{\delta}))(1 + n\delta + n\hat{\delta})}{(1 + n\hat{\delta})(1 + n\delta + n\hat{\delta}) - \delta} - 1 \right) \hat{\Delta} x.
\end{equation}

The purpose of this exercise is to construct a counterfactual that allows us to evaluate the GMM without assuming any potential increase in system-wide liquidity.

Again, Table~\ref{tab:n} shows a significant reduction in the profitability of sandwich attacks and a high fraction of cases in which these profits disappear.

\begin{table}[H]
\centering
\begin{tabular}{lcc}
\toprule
\textbf{Algorithm} & \textbf{Attackers' profits} & \textbf{Attacks with profits$<$0} \\
\midrule
Zeromev     & \$338.70m & 4.1 \% \\
CPMM   & \$333.70m & 3.30\% \\
GMM (n=2) & \$44.26m  & 78.22\% \\
GMM (n=3) & \$27.09m  & 84.01\% \\
GMM (n=4) & \$19.58m  & 87.16\% \\
GMM (n=5) & \$15.37m  & 89.20\% \\
\bottomrule
\end{tabular}
\caption{Sandwiches reported by Zeromev and simulations for 2023: attacker's profits and percentage of attacks with negative profits for different values of the number~$n$ of equal-sized AMMs into which the Uniswap V2 is split.}
\label{tab:n}
\end{table}

Finally, we conclude our analysis by using our dataset to compare the CPMM and the GMM in terms of impermanent losses. This is a complex task, as it requires isolating the effects of various events that influence AMM reserves over time, primarily liquidity injections, withdrawals, and fee revenue. To simplify the computation while still obtaining a realistic estimate, we rely on Equations~\eqref{eq:ILCPMM} and~\eqref{eq:ILGMM}. These formulae allow us to calculate impermanent losses as a function of the change in the relative price of the crypto assets in the AMM’s liquidity pool between two points in time, and the share of the AMM’s reserves relative to the global reserves across all AMMs considered in the GMM. The main caveat is that Equation~\eqref{eq:ILGMM} assumes only two AMMs are included in the GMM formula, and that a single trade moves the AMM’s reserves to the point where its marginal price matches the new relative price. Allowing for multiple intermediate trades can only increase the value of the final portfolio of the AMM, thereby reducing impermanent losses. Similarly, incorporating more AMMs into the formula can only improve the performance of the GMM. In this sense, our assumptions provide a conservative estimate of the potential gains from adopting the GMM in place of the CPMM.

Equation~\eqref{fig:GMM_IL} requires the initial and final relative prices of the crypto assets as input. For each pair, we compute these prices by identifying, in our sequence of transactions, the first and last time the pair was traded in Uniswap V2.\footnote{In this process, we drop pairs Pairs without reliable price data in US\$, pairs with less than two transactions and those with values so close to zero that rounding errors suggest dropping them from the analysis. These reduces the number of pairs from 165,566 to 124,898.}
The total impermanent losses computed for the CPMM algorithm using Equation~\eqref{eq:ILCPMM} amount to \$615.3 million. Equation~\eqref{eq:ILGMM} also requires the parameter $\alpha$, which denotes the share of Uniswap's reserves relative to the global reserves considered under the GMM. As previously discussed, accurately tracking the reserves of other AMMs trading the same token pairs is a complex task. For simplicity, we select illustrative values of $\alpha$.\footnote{Recall that Equation~\eqref{eq:ILGMM} requires $\alpha\leq 0.5$.} This corresponds to a counterfactual in which Uniswap V2 uses the GMM algorithm and incorporates the reserves of another AMM whose size is $\frac{1 - \alpha}{\alpha}$ times that of Uniswap V2.

Finally, we distinguish between pairs with low and high price volatility. The later when the relative price from the beginning to the end of 2023 varies more than 10 times. 

\begin{table}[h!]
\centering
\begin{tabular}{c| cc |cc}
\toprule
\textbf{$\alpha$} 
& \multicolumn{2}{c|}{\textbf{Low Volatility }} 
& \multicolumn{2}{c}{\textbf{High Volatility }} \\
& \textbf{$IL_{CPMM}-IL_{GMM}$} & \textbf{\% of $IL_{CPMM}$} 
& \textbf{$IL_{CPMM}-IL_{GMM}$} & \textbf{\%  of $IL_{CPMM}$} \\
\midrule
$0.01$  & \$12.5m & 98.4\%  & \$136.6m & 22.7\% \\
$0.05$  & \$12.0m & 94.5\%  & \$93.8m  & 15.6\% \\
$0.1$   & \$11.3m & 89.0\%  & \$76.9m  & 12.8\% \\
$0.25$  & \$9.3m  & 73.2\%  & \$53.8m  & 8.9\%  \\
$0.5$   & \$6.1m  & 48.0\%  & \$33.3m  & 5.5\%  \\
\bottomrule
\end{tabular}
\caption{Reduction in impermanent losses for different values of $\alpha$ under the GMM. Percentages are relative to the CPMM impermanent losses of \$12.7m (low volatility) and \$602.6m (high volatility) under the CPMM.}
\label{tab:GMM_IL}
\end{table}

Table~\ref{tab:GMM_IL} shows that the reduction in impermanent losses is substantial and largely driven by volatile pairs. These cases likely correspond to pairs in which one of the crypto assets failed, causing its relative price to collapse to near zero. As a result, liquidity providers who had deposited funds in these pools lost almost all their capital. To shed light on the magnitude of these price shifts, Table~\ref{tab:volatility_ranges} reports the fraction of token pairs that experienced price changes by factors of 10, 100, and 1000.

\begin{table}[ht]
\centering
\begin{tabular}{ccc}
  \toprule
 & Low Volatility Pairs & High Volatility Pairs\\
  \midrule
$\lambda=10$ & 87.5 \%& 12.5 \%\\ 
  $\lambda=100$ & 92.1 \%& 7.9 \%\\ 
  $\lambda=1000$ & 93.1 \%& 6.9 \%\\ 
   \bottomrule
\end{tabular}
\caption{Percentage of pairs with price variations greater than a factor of $\lambda$, for different values of $T$. The total number of pairs is 124,898.} 
\label{tab:volatility_ranges}
\end{table}

%% file: sections/Implementation.tex
\label{sec:impl}

In this section, we overview the technical feasibility of our proposed algorithm, the Generalized Market Maker (GMM), and provide some statistics on the cost of using the algorithm in terms of gas consumption on the Ethereum blockchain.

The implementation of the GMM in Solidity, the coding lenguage of Ethereum, is publicly available in \url{https://github.com/JonathanJimenezMunoz/GMM/tree/main}\footnote{We thank Jonathan Jimenez Muñoz for developing this implementation.} The design is based on the architecture of Uniswap V2, the most widely used implementation of CPMM, and modifies only the components necessary to allow the algorithm to account for global reserves. Thus, the implementation inherits the robust and battle-tested foundation of Uniswap V2. Besides, the modular architecture of Uniswap V2 facilitates maintainability, integration with existing infrastructure and future updates.

The computational cost of the GMM, measured in gas consumption (the unit of computational effort on the Ethereum network) together with its cost in dollars, is reported in Table~\ref{tab:gas_costs}. We compare the original CPMM implementation with the modified GMM version for different numbers of external AMMs included in the global reserve calculation. As expected, gas usage increases with the number of external references, since each additional AMM requires an extra data retrieval and computation step. However, the increase in gas cost is moderate and consistent with the additional informational complexity introduced by the GMM. These results suggest that the GMM can be deployed in practice with manageable increases in gas consumption. 

\begin{table}[H]
\centering
\begin{tabular}{lrr}
\toprule
\textbf{Algorithm} & \textbf{Gas Consumption} & \textbf{Cost} \\
\midrule
CPMM & 102{,}438 & \$1.61 \\
GMM (1 other AMM) & 115{,}676 & \$1.82 \\
GMM (2 other AMMs) & 120{,}619 & \$1.90 \\
GMM (3 other AMMs) & 125{,}091 & \$1.97 \\
GMM (4 other AMMs) & 129{,}492 & \$2.04 \\
GMM (5 other AMMs) & 133{,}916 & \$2.11 \\
GMM (6 other AMMs) & 138{,}317 & \$2.18 \\
\bottomrule
\end{tabular}
\caption{Gas consumption and cost for CPMM and GMM with varying numbers of other AMMs. Prices in dollars using a gas cost of 5.764 Gwei and ETH price of \$2,726.29, as of May 27th, 2025.
} 
\label{tab:gas_costs}
\end{table}

%% file: sections/conclusion.tex
\label{sec:con}

In this paper, we have analyzed how to improve an existing pricing algorithm used in automated market makers by leveraging economic analysis. Our focus has been on three key metrics: arbitrage costs, MEV sandwich attacks, and impermanent losses.

However, our investigation leaves aside some important aspects that warrant further research. The first is the performance of our algorithm, the GMM, when adopted by only a subset of market participants. Understanding this is crucial to determining whether there are incentives to transition toward a market structure in which all AMMs adopt the GMM. Additionally, it is relevant to assess whether the GMM benefits smaller players by improving their competitiveness or, conversely, reinforces the market power of larger players. Furthermore, partial adoption of the GMM introduces new trade-offs, as arbitrage opportunities are not fully eliminated unless the transition is complete.

The second aspect is the role of fees. Existing AMM algorithms, such as the CPMM, use fees to compensate liquidity providers and mitigate impermanent loss. The GMM, however, increases the reserves within the AMM without requiring fees. Additionally, by reducing impermanent losses, the GMM allows for lower fees without compromising the compensation of liquidity providers. A deeper analysis of these factors would provide a more comprehensive understanding of how the efficiency gains from the GMM are distributed between traders and liquidity providers.

Finally, our work introduces a new field of economic analysis: the design of smart contracts. This approach is related to the methodology used by \cite{sonmez2024} in designing market mechanisms for kidney exchange and school allocation, among other applications. Instead of optimizing based on traditional criteria such as revenue or surplus—subject to constraints like incentive compatibility—we adapt an existing practical implementation, the CPMM algorithm, to incorporate a meaningful improvement: better information gathering. This adaptation is constrained by desirable properties, such as ensuring that neither the AMM nor traders are worse off in a well-defined sense. We believe that this market design perspective on smart contract development can lead to further valuable applications in decentralized finance and beyond.